\documentclass[english,prb,aps,amsmath,amssymb,reprint,superscriptaddress,showkeys]{revtex4-2}
\usepackage{mathptmx}
\usepackage[T1]{fontenc}
\usepackage[utf8]{inputenc}
\usepackage{color}
\usepackage{babel}
\usepackage{float}
\usepackage{textcomp}
\usepackage{amsmath}
\usepackage{amssymb}
\usepackage{graphicx}
\usepackage{wasysym}
\usepackage[unicode=true,
 bookmarks=true,bookmarksnumbered=false,bookmarksopen=false,
 breaklinks=true,pdfborder={0 0 0},pdfborderstyle={},backref=false,colorlinks=true]
 {hyperref}
\hypersetup{pdftitle={Hydrogen-related reactions and carrier traps in solar silicon from first principles},
 pdfauthor={José Coutinho},
 allcolors=blue}

\makeatletter

\providecommand{\tabularnewline}{\\}

\usepackage{orcidlink}

\makeatother

\begin{document}
\title{Hydrogen reactions with dopants and impurities in solar silicon from
first principles}
\author{José Coutinho\orcidlink{0000-0003-0280-366X}}
\affiliation{i3N and Department of Physics, University of Aveiro, Campus Santiago,
3810-193 Aveiro, Portugal}
\email{jose.coutinho@ua.pt}

\author{Diana Gomes\orcidlink{0000-0001-8263-8523}}
\affiliation{i3N and Department of Physics, University of Aveiro, Campus Santiago,
3810-193 Aveiro, Portugal}
\author{Vitor J. B. Torres\orcidlink{0000-0003-3795-6818}}
\affiliation{i3N and Department of Physics, University of Aveiro, Campus Santiago,
3810-193 Aveiro, Portugal}
\author{Tarek O. Abdul Fattah\orcidlink{0000-0001-8734-6706}}
\affiliation{Photon Science Institute and Department of Electrical and Electronic
Engineering, The University of Manchester, Manchester M13 9PL, United
Kingdom}
\author{Vladimir P. Markevich\orcidlink{0000-0002-2503-6144}}
\affiliation{Photon Science Institute and Department of Electrical and Electronic
Engineering, The University of Manchester, Manchester M13 9PL, United
Kingdom}
\author{Anthony R. Peaker\orcidlink{0000-0001-7667-4624}}
\affiliation{Photon Science Institute and Department of Electrical and Electronic
Engineering, The University of Manchester, Manchester M13 9PL, United
Kingdom}
\begin{abstract}
We present a theoretical account of some of the most likely hydrogen-related
reactions with impurities in n-type and p-type solar-grade silicon.
These include reactions with dopants and carbon, which are relevant
in the context of life-time degradation of silicon solar cells, most
notably of light and elevated temperature degradation (LeTID) of the
cells. Among the problems addressed, we highlight a comparative study
of acceptor-enhanced dissociation of hydrogen molecules in B- and
Ga-doped material, their subsequent reaction steps toward formation
of acceptor-hydrogen pairs, the proposal of mechanisms which explain
the observed kinetics of photo-/carrier-induced dissociation of PH
and CH pairs in n-type Si, analysis of reactions involving direct
interactions between molecules with P and C, and the assignment of
several electron and hole traps with detailed atomistic- and wavefunction-resolved
models.

\noindent \emph{Pre-print published in Solar RRL 8, 2300639 (2024)}

\noindent DOI:\href{https://doi.org/10.1002/solr.202300639}{10.1002/solr.202300639}
\end{abstract}
\maketitle

\section{Introduction}

Over the last two decades, a combination of cost reduction with improvement
in material quality, changed the list of concerns regarding the various
sources of bulk life-time degradation of solar silicon. For instance,
while transition metal contamination became avoidable, the boron-oxygen
light induced degradation gained relevance \citep{schmidt2004,lindroos2016,niewelt2017},
justifying a first push toward replacing boron (a long-standing dopant
of choice), by alternative dopants \citep{murphy2019,chen2021,moeller2013,cho2016,Fattah2023a}.
According to the International Technology Roadmap for Photovoltaic
\citep{itrpv2023}, Ga-doped p-type Cz-Si already dominates the market
share in 2023, and will remain the dominant material within the next
10 years.

A second and more recent reason for exploring alternative dopants
other than boron, has been Light- and elevated-Temperature-Induced
Degradation (LeTID) of solar Si \citep{ramspeck2012,bredemeier2019}.
This is a crucially important degradation mechanism, as modules show
a substantial decrease of conversion efficiency (up to 16\% relative)
during their operating life. It is now recognized that LeTID can affect
cells made of different Si materials, including p- and n-type \citep{chen2017,chen2018,lin2021,grant2021},
although it is not clear if the source of the problem is the same
in all cases.

While degradation in the field takes years to manifest, LeTID can
be accelerated by increasing the operating temperature to $\sim70$-100~°C
with illumination of $\sim1$~sun (1~kW/m$^{2}$), or more usually
in lab tests by passing a current to inject minority carriers equivalent
to the population produced during normal operation. This is sometimes
referred to as carrier induced degradation (CID), and produces very
similar results to illumination. Many variants of accelerated LeTID/CID
appear in the literature ranging from high intensity laser irradiation
to dark annealing. It seems nevertheless that degradation and recovery
take place simultaneously at different rates with thermal activation
energies close to 1~eV \citep{chen2020a}. This may go some way to
explain the differences in observed behavior, but in an attempt to
standardize rapid LeTID testing, procedures for commercial applications
have been drawn up \emph{e.g.} SEMI~PV93 and IEC~61215. These testing
methods also include methodologies that minimize interference from
iron-related and boron-oxygen light induced degradation effects.

Importantly, there is a very substantial literature on LeTID which
relates the magnitude of the effect to the fast-firing step in the
fabrication process. Although there are wide variations in the results,
it now seems certain that hydrogen is implicated (see for instance
Refs.~\citep{chen2020a,chen2020b} and references therein).

Molecular hydrogen is well know to form upon quenching p-type and
n-type Si in contact with a hydrogen source at high temperature \citep{pritchard1998,pritchard1999,stavola2003}.
The molecules constitute the main stock of hydrogen in the Si of \emph{as-fired}
solar cells, and originates from the hydrogen-rich passivating oxide/nitride
stack \citep{sheoran2008}. In pristine Si they become mobile just
above room temperature. However, in O-rich material they are lightly
bound to interstitial oxygen atoms and one needs to raise the temperature
over 70~°C to initiate molecular motion \citep{markevich1998}, and
possibly, to induce LeTID.

Several models for LeTID have been hypothesized, but as yet, a detailed
mechanism or mechanisms remain unclear. We recently proposed that
a possible culprit for the non-radiative recombination activity behind
LeTID in boron-doped Si is a complex made of boron and two hydrogen
atoms \citep{deguzman2021}. Accordingly, BH$_{2}$ is a byproduct
of the reaction between a non-equilibrium population of H$_{2}$ molecules
with boron, when the molecules become mobile above room temperature.
From first-principles calculations, we found that the first and second
steps toward the right of the reaction,

\begin{equation}
\textrm{H}_{2}+2\textrm{B}^{-}+2\textrm{h}^{+}\rightarrow\textrm{BH}_{2}^{+}+\textrm{B}^{-}\rightarrow2\textrm{BH},\label{reaction1}
\end{equation}
were limited by activation barriers of 1.1~eV and 0.8~eV \citep{coutinho2023}.
These are considerably lower than the barrier for H$_{2}$ dissociation
in pristine Si (1.6~eV \citep{gomes2022}), and hence boron was claimed
to act as a catalyst for the breaking of H$_{2}$, and facilitate
BH pair formation.

The calculated value of the energy barrier for the first reaction
in sequence \ref{reaction1} accounts well for the activation energy
of the degradation stage of 1.08~eV determined for B-doped multi
crystalline Si \citep{vargas2019}. It is also in line with recent
findings which show that LeTID develops concurrently with the early
stages of BH formation \citep{hammann2021,hammann2023}.

It has been argued in Ref.~\citep{fattah2023b} that GaH$_{2}$ complexes
are effective-mass-like shallow donors, and unlike BH$_{2}$, they
should not act as recombination centers. It is therefore hard to understand
how a GaH$_{2}$ analogue of the BH$_{2}$ complex could explain the
observation of LeTID in cells made from Ga-doped substrates \citep{grant2021,kwapil2021,post2022}.
Similar arguments apply to the degradation observed in n-type based
cells (although in this case the effect is weaker and slower) \citep{chen2018,kang2019,chen2020b}.
Possibilities could involve contamination of the materials with boron,
or alternatively that H$_{2}$ dissociation is just the first step
for the formation of a mix of different hydrogen-related electrically
active centers (among which we have BH$_{2}$).

Presently, passivated emitter and rear cell (PERC) architecture dominates
the photovoltaic market \citep{green2015}. Front and back passivation
layers make these devices particularly vulnerable to LeTID. Solar
cells using new concepts like n-type based tunneling oxide passivated
contacts (TOPCon) and silicon heterojunction (SHJ) cells, although
show improved lifetime stability, also seem to be affected by LeTID
\citep{chen2021}.

It is therefore of uttermost importance to understand the reactions
involving hydrogen in n-type Si, including the effect of minority
carriers, and ideally using state-of-the-art quantum mechanical calculations.
We cannot provide an extensive review of previous work on hydrogen
in Si. For that we direct the reader for instance to Refs.~\citep{nickel1999,estreicher2015,gomes2022}
and references therein. It is however useful to summarize some properties
which are especially relevant in the present context.

Atomic H in Si is an amphoteric element --- it is an acceptor and
a donor in n-type and p-type Si, respectively \citep{nielsen1999,herring2001,nielsen2002}.
H$^{+}$ sits at the center of a Si-Si bond, while H$^{-}$ occupies
a tetrahedral interstitial cage \citep{deak1988,vandewalle1989}.
The neutral state is metastable and disproportionates into ionic species,
$\textrm{H}^{0}\rightarrow x\textrm{H}^{+}+(1-x)\textrm{H}^{-}$,
with $x$ depending on the Fermi level with respect to the negative-$U$
$(-/+)$ transition \citep{vandewalle1989,johnson1994}. Isolated
hydrogen is very reactive and mobile, making it virtually inexistent
under equilibrium conditions at room temperature. However, it can
show up transiently upon changes of thermodynamic or excitation conditions
(e.g. release of H from H-related complexes upon capture of photogenerated
carriers). The fractional populations of charge states of free H as
a function of temperature has been described by Sun \emph{et~al.}
\citep{sun2015,sun2021} using a \emph{general occupancy ratio} model,
which incorporates not only information regarding the transition levels
of the defects, but also their capture cross sections for free carriers.

Due to its high mobility (the diffusion length of $\textrm{H}^{+}$
is about $1.4\:\mu\textrm{m}$ at room temperature \citep{gorelkinskii1996}),
hydrogen interacts with other available lattice defects, most effectively
with defects which have an opposite charge \citep{gorelkinskii1996,langpape1997,nielsen1999}.
The problem of diffusivity of H atoms in differently doped Si crystals
has been considered recently in Ref.~\citep{gomes2022}. Reactions
between hydrogen and group-III elements in Si were also addressed
theoretically using hybrid density functional theory \citep{coutinho2023}.
However, a comparable study of interactions between hydrogen and group-V
elements is not available, and as far as we are aware, there are no
reports either demonstrating or ruling out the existence of direct
interactions between H$_{2}$ and donors in Si. A clarification of
this issue would be welcome in the context of solar Si.

Hydrogen passivation of group-V donors in Si is a well documented
topic \citep{johnson1986,bergman1988,chang1988,estreicher1989,zhu1990,tavendale1990,fukata2007}.
Phosphorus-H defects show a $\equiv\!\textrm{P}\;\;\textrm{Si-H}_{\textrm{AB}}$
geometry, where the H atom binds to Si, oppositely to a broken P-Si
bond, leaving the P atom three-fold coordinated \citep{chang1988,estreicher1989,denteneer1990}.
This location is commonly referred to as \emph{anti-bonding} (AB)
site, and contrasts with the\emph{ bond-center} (BC) site found for
the acceptor-H pairs.

Donor-H complexes anneal out around $T\sim80\textrm{-}100$~ºC, among
which PH is the most stable, with dissociation barriers estimated
in the range 1.1-1.2~eV \citep{zhu1990,pearton1991}. Importantly,
in the presence of minority carriers, donor-H pairs become unstable,
even below room temperature \citep{seager1990,johnson1992b,seager1993}.
The effect has been explained by two alternative views: a \emph{dissociative
model} \citep{johnson1992}, where thermal fluctuations promote partial
dissociation of PH into close P$^{+}$-H$^{-}$ pairs, enabling hole
capture by hydrogen, and subsequent escape of H$^{0}$ or H$^{+}$
from the Coulomb field of the donor. After that, and in the absence
of minority carriers (zero bias and darkness), the measured recovery
kinetics of resistivity has an activation barrier of 0.7~eV, interpreted
as the upper limit for the migration energy of H$^{-}$ before reformation
of the pairs \citep{johnson1992,herring2001}.

The other view follows a \emph{transformative model} \citep{estreicher1991,estreicher1994},
where upon hole capture the $\equiv\!\textrm{P}\;\;\textrm{Si-H}_{\textrm{AB}}$
ground state quickly converts into a more stable positively charged
$\equiv\!\textrm{P}^{+}\;\;\textrm{H}_{\textrm{BC}}\textrm{-Si}$
state, where H jumps into the center of the P-Si bond. The observed
0.7~eV barrier of the passivation recovery kinetics measured in Ref.~\citep{johnson1992},
is in this case attributed to the activation energy of the reverse
jump of H, from BC to the AB site in the neutral state after electron
capture. However, this picture is only possible if PH has a donor
state, a property that has never been demonstrated (nor refuted) experimentally
or theoretically.

Other omnipresent impurities in solar Si are carbon and oxygen. The
available experimental and theoretical results indicate that OH pairs
are weakly-bound and not stable at room temperature \citep{estreicher1990,nielsen1997,ramamoorthy1998,nielsen1999,capaz1999,nielsen2002},
and therefore are not addressed here. Oxygen-H$_{2}$ pairs are not
electrically active, and their relevance (especially within the context
of LeTID) seems to be that of stabilizing the molecules, \emph{i.e.},
hindering molecular motion at room temperature \citep{markevich1998}.

Carbon, on the other hand, is able to form stable and electrically
active centers upon capture of hydrogen atoms. This paper extends
our understanding regarding the formation and properties of CH$_{n}$
complexes. We focus on reactions involving substitutional carbon and
hydrogen under typical conditions of solar cell fabrication and operation.
In this case, already known defects are the CH pair and CH$_{2}$
complex. The latter, also referred to as CH$_{2}^{*}$ \citep{suezawa1999,markevich2001,hourahine2001,estreicher2012},
is stable up to about 250~ºC and it is electrically inert.

Regarding the CH pair, the picture is not so consensual. According
to combined first-principles calculations and Laplace deep level transient
spectroscopy (Laplace-DLTS) \citep{andersen2002}, the most stable
CH defect displays a $\equiv\textrm{C-H}_{\textrm{BC}}\;\;\textrm{Si}\equiv$
geometry (possessing an unsaturated Si radical). The defect was connected
to donor and acceptor transitions respectively at $E_{\textrm{v}}$+0.33~eV
and $E_{\textrm{c}}-0.16$~eV, and in n-type Si it is stable up to
just above room temperature in darkness \citep{andersen2002}. However,
it rapidly disappears under above-band-gap illumination \citep{kamiura1997}.
In p-type Si the CH pair is more stable and anneals out just above
100~ºC \citep{kamiura1995}. The above effects could be important
in the context of light-induced reactions in solar silicon, but the
mechanisms remain unexplored theoretically.

The above picture for the CH complex, including the structure, location
of transition levels, as well as the number of charge states, are
also under dispute. Recent capacitance and depth profile measurements
led to an alternative view, where several CH complexes could form
in wet-etched and plasma-treated material. While there is an agreement
regarding the origin of the $E_{\textrm{c}}-0.16$~eV acceptor transition
(the $\equiv\textrm{C-H}_{\textrm{BC}}\;\;\textrm{Si}\equiv$ defect),
the $E_{\textrm{v}}$+0.33~eV level was assigned to an acceptor transition
of a C-H complex with a different geometry. Additionally, first and
second acceptor levels at $E_{\textrm{c}}-0.51$~eV and $E_{\textrm{c}}-0.06$~eV
were assigned to a H$_{\textrm{AB}}$-C~~Si$\equiv$ defect, and
a complex with more than one H atom was assigned to an electron trap
at $E_{\textrm{c}}-0.14$~eV \citep{kolkovsky2018}. In that respect,
we hope to contribute to the clarification of the matter, especially
considering that some of these centers may form transiently, during
the relocation of H upon heating or illumination, and possibly lower
the minority carrier lifetime in both p-type and n-type Si.

The paper is organized as follows: Section.~\ref{sec:AH} describes
a comparative study of hydrogen reactions (atomic and molecular) with
boron and gallium acceptors in p-type Si. In Secs.~\ref{sec:PH}
and \ref{sec:PH2} we investigate analogous reactions with phosphorus
(n-type Si), with emphasis on the minority-carrier-enhanced dissociation
of PH. In Secs.~\ref{sec:CH} and \ref{sec:CH2} we revisit and extend
previous calculations of CH$_{n}$ complexes with $n\leq3$. Finally,
we lay our conclusions in Sec.~\ref{sec:conclusions}.

\section{Dissociation of molecular hydrogen upon reaction with acceptors\label{sec:AH}}

H$_{2}$ molecules occupy tetrahedral interstitial cages of the Si
lattice. Inspection of the band structure of a supercell with the
molecule indicates that it has a clean band gap, \emph{i.e.} cannot
trap carriers, irrespectively of its orientation and location within
the accessible volume. A fully occupied Kohn-Sham state in the lower
half of the gap, which could lead to carrier trapping, appears only
after partial dissociation of the molecule into a H$_{\textrm{BC}}^{+}$-H$_{\textrm{AB}}^{-}$
pair. This state is achieved upon collision of the molecule with a
Si-Si bond, after surmounting an energy barrier of 1.6~eV, and that
figure was assigned to the dissociation barrier of isolated H$_{2}$
\citep{gomes2022}.

More recently, we have argued that boron can act as a catalyst for
H$_{2}$ dissociation by lowering the above barrier to 1.1~eV \citep{coutinho2023}.
The mechanism involves H$_{2}$ hitting a Si-B bond and subsequent
formation of a metastable state comprising H$^{-}$ next to a BH pair
(BH-H$^{-}$). Several processes were hypothesized to follow (including
hole capture by the hydride anion), but importantly, the $\textrm{H}_{2}+\textrm{B}^{-}\rightarrow\textrm{BH}\textrm{-}\textrm{H}^{-}$
step was deemed critical for triggering LeTID in B-doped Si.

Given the controversial results on LeTID in cells based on Ga-doped
Si, we have performed a comparative study of H-reactions with B and
Ga in Si. Some results already reported include the binding and dissociation
energies of BH and GaH pairs. The binding energy was found from the
reaction energy $E_{\textrm{b}}=\Delta E_{\textrm{R}}=E_{\textrm{fs}}-E_{\textrm{is}}$
across $X\textrm{H}\rightarrow X^{-}+\textrm{H}^{+}$ (with $X=\{\textrm{B},\,\textrm{Ga}\}$),
where $E_{\textrm{is}}$ and $E_{\textrm{fs}}$ stand for the energy
of initial (reactants) and final (products) states, respectively.
The dissociation energy of a defect complex is the overall barrier
that the system must surmount, starting from its ground state, and
reach the dissociated state (uncorrelated constituents, for instance
infinitely separated $\textrm{B}^{-}$ and $\textrm{H}^{+}$). This
is obtained from the activation energy $E_{\textrm{d}}=\Delta E_{\textrm{A}}=E_{\textrm{ts}}-E_{\textrm{is}}$
of the above reaction, where $E_{\textrm{ts}}$ is the energy of the
transition state, all along the minimum-energy path between reactants
and products evaluated using the NEB method.

In principle, the dissociation mechanism involves an infinite sequence
of $\textrm{H}^{+}$ jumps away from $\textrm{B}^{-}$. To make the
problem tractable, we calculated the barriers of $\textrm{H}^{+}$
jumps from first and second neighboring sites (with respect to the
$\textrm{B}^{-}$ ion), and from sites where $\textrm{H}^{+}$ is
infinitely separated from $\textrm{B}^{-}$. First and second jumps
of H away from B and Ga have transition-state energies that are clearly
below that of saddle point of $\textrm{H}^{+}$ migration at a remote
location from $\textrm{B}^{-}$ \citep{coutinho2023,fattah2023b}.
Assuming that the energy barriers of $\textrm{H}^{+}$ jumps from
sites farther than third neighbors (from $X^{-}$) are identical to
that of isolated $\textrm{H}^{+}$ (calculated as $E_{\textrm{m}}=0.42$~eV),
we conclude that the dissociation energy of the pairs is $E_{\textrm{d}}=E_{\textrm{b}}+E_{\textrm{m}}$.

The results are shown in Table~\ref{tab1} (the boron-related data
is reproduced from Ref.~\citep{coutinho2023}). For BH, the calculated
binding and dissociation energies are $E_{\textrm{b}}=0.76$~eV and
$E_{\textrm{d}}=E_{\textrm{b}}+E_{\textrm{m}}=1.18$~eV. They are
relatively lower than the $E_{\textrm{b}}=0.92$~eV and $E_{\textrm{d}}=1.34$~eV
analogues for GaH. These results agree well with those derived from
capacitance-voltage measurements by Zundel and Weber \citep{zundel1989},
who reported $E_{\textrm{d}}=1.28$~eV and 1.40~eV for BH and GaH,
respectively.

The comparison between experimental and calculated activation energies
of defect reactions should be done with care. Usually, several defect
reactions, including interactions with carriers, take place in parallel
during experiments designed to follow the loss or growth of a specific
complex. For instance, reliable values of dissociation energies of
$X$H pairs, free of some back-reaction effects, could only be measured
in the depletion region of reverse-biased diodes \citep{zundel1989}.
The excellent agreement between those results and our calculations
is reassuring regarding the interpretation of the data.

\noindent 
\begin{table}
\caption{\label{tab1}Calculated reaction and activation energies ($\Delta E_{\textrm{R}}$
and $\Delta E_{\textrm{A}}$, respectively) for formation of acceptor-hydrogen
complexes in p-type silicon. $\Delta E_{\textrm{R}}=E_{\textrm{fs}}-E_{\textrm{is}}$
and $\Delta E_{\textrm{A}}=E_{\textrm{ts}}-E_{\textrm{is}}$, where
subscripts \textquoteleft is\textquoteright , \textquoteleft ts\textquoteright{}
and \textquoteleft fs\textquoteright{} stand for initial, transition
and final states of the reaction on the leftmost column, respectively.
All values are in eV.}

\begin{tabular}{lllr@{\extracolsep{0pt}.}lr@{\extracolsep{0pt}.}lr@{\extracolsep{0pt}.}lr@{\extracolsep{0pt}.}lr@{\extracolsep{0pt}.}lr@{\extracolsep{0pt}.}lr@{\extracolsep{0pt}.}l}
\hline 
Acceptor ($X$) &  &  & \multicolumn{4}{c}{Boron} & \multicolumn{2}{c}{} & \multicolumn{2}{c}{} & \multicolumn{2}{c}{} & \multicolumn{4}{c}{Gallium}\tabularnewline
Reaction &  &  & \multicolumn{2}{c}{$\Delta E_{\textrm{R}}$} & \multicolumn{2}{c}{$\Delta E_{\textrm{A}}$} & \multicolumn{2}{c}{} & \multicolumn{2}{c}{} & \multicolumn{2}{c}{} & \multicolumn{2}{c}{$\Delta E_{\textrm{R}}$} & \multicolumn{2}{c}{$\Delta E_{\textrm{A}}$}\tabularnewline
\hline 
$X^{-}+\textrm{H}^{+}\rightarrow X\textrm{H}$ &  &  & \multicolumn{2}{c}{$-0.76$} & \multicolumn{2}{c}{$0.42$} & \multicolumn{2}{c}{} & \multicolumn{2}{c}{} & \multicolumn{2}{c}{} & \multicolumn{2}{c}{$-0.92$} & \multicolumn{2}{c}{$0.42$}\tabularnewline
$2X\textrm{H}\rightarrow X^{-}+X\textrm{H}_{2}^{+}$ &  &  & \multicolumn{2}{c}{$+0.36$} & \multicolumn{2}{c}{$1.18$} & \multicolumn{2}{c}{} & \multicolumn{2}{c}{} & \multicolumn{2}{c}{} & \multicolumn{2}{c}{$+0.43$} & \multicolumn{2}{c}{$1.34$}\tabularnewline
$X\textrm{H}+\textrm{H}^{+}\rightarrow X\textrm{H}_{2}^{+}$ &  &  & \multicolumn{2}{c}{$-0.40$} & \multicolumn{2}{c}{$0.42$} & \multicolumn{2}{c}{} & \multicolumn{2}{c}{} & \multicolumn{2}{c}{} & \multicolumn{2}{c}{$-0.48$} & \multicolumn{2}{c}{$0.42$}\tabularnewline
$\textrm{H}_{2}+X^{-}\rightarrow X\textrm{H}_{2}^{-}$ &  &  & \multicolumn{2}{c}{$-0.31$} & \multicolumn{2}{c}{$1.1$} & \multicolumn{2}{c}{} & \multicolumn{2}{c}{} & \multicolumn{2}{c}{} & \multicolumn{2}{c}{$+0.14$} & \multicolumn{2}{c}{$1.4$}\tabularnewline
$\textrm{H}_{2}+X^{-}+2\textrm{h}^{+}\rightarrow X\textrm{H}_{2}^{+}$ &  &  & \multicolumn{2}{c}{$-0.83$} & 1&10 & \multicolumn{2}{c}{} & \multicolumn{2}{c}{} & \multicolumn{2}{c}{} & \multicolumn{2}{c}{$-1.07$} & 1&05\tabularnewline
$\textrm{H}_{2}+2X^{-}+2\textrm{h}^{+}\rightarrow2X\textrm{H}$ &  &  & \multicolumn{2}{c}{$-1.19$} & 1&10 & \multicolumn{2}{c}{} & \multicolumn{2}{c}{} & \multicolumn{2}{c}{} & \multicolumn{2}{c}{$-1.50$} & 1&05\tabularnewline
\hline 
\end{tabular}
\end{table}

From a comprehensive exploration of the potential energy surface involving
the motion of H$^{+}$ next to B$^{-}$ and Ga$^{-}$ we could arrive
at the reaction and dissociation energies reported in the topmost
three rows of Table~\ref{tab1} \citep{fattah2023b,coutinho2023}.
Disproportionation of two $X$H defects into $X\textrm{H}_{2}^{+}$
and $X^{-}$ (second row) is endothermic, and therefore $X\textrm{H}_{2}^{+}$
is metastable with respect to $X$H formation. Along such $\textrm{H}^{+}$
exchange reaction, the total energy is always lower than that of the
intermediate $X\textrm{H}+X^{-}+\textrm{H}^{+}$ state plus the migration
barrier of H$^{+}$. Hence, the activation energy for $2X\textrm{H}\rightarrow X^{-}+X\textrm{H}_{2}^{+}$
corresponds to the dissociation barrier of $X$H.

Despite being metastable with respect to $X$H formation, $X\textrm{H}_{2}^{+}$
complexes can form upon capture of H$^{+}$ by $X$H pairs in the
presence of large concentrations of atomic hydrogen \citep{deguzman2021,fattah2023b}.
The reaction (third row of Table~\ref{tab1}) is energetically favorable
($\Delta E_{\textrm{R}}=-0.40$~eV and $-0.48$~eV for $X=\textrm{B}$
and Ga, respectively). The very last barrier before formation of $X\textrm{H}_{2}^{+}$,
which involves a jump of H$^{+}$ into the center of a $X$-Si bond
of the $X$H pair ($X\textrm{H}\textrm{-}\textrm{H}^{+}\rightarrow X\textrm{H}_{2}^{+}$),
was calculated as 0.46~eV and 0.36~eV for $X=\textrm{B}$ and Ga,
respectively. Hence, formation of $X\textrm{H}_{2}^{+}$ via capture
of $\textrm{H}^{+}$ by $X$H in $X$-doped Si is anticipated to show
an activation energy similar to the migration barrier of H$^{+}$.

\begin{figure*}
\includegraphics[width=18cm]{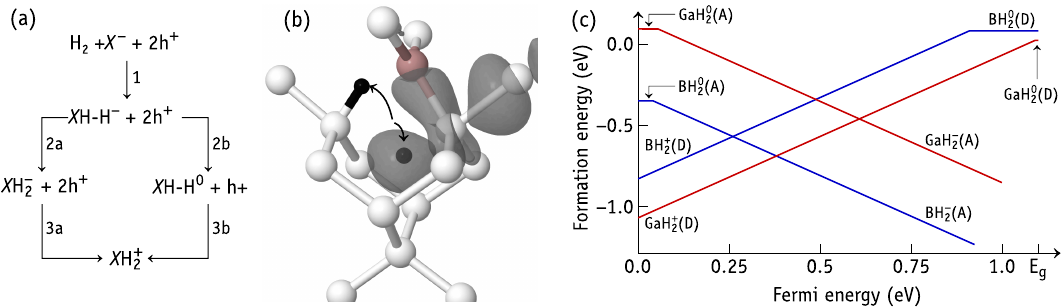}

\caption{\label{fig1}(a) Proposed mechanism for dissociation of $\textrm{H}_{2}$
molecules next to acceptor impurities in p-type silicon and subsequent
formation of $X\textrm{H}_{2}^{+}$ complex. (b) Atomistic structure
of the $\textrm{GaH}\textrm{-H}^{-}$ complex obtained after step
1, along with the electron density associated with the highest occupied
gap state (gray isosurface), which is likely to trap holes. Si, Ga
and H are shown in white, pink and black balls, respectively. The
arrows depict an approximate trajectory of H atoms upon dissociation.
(c) formation energy diagram of BH$_{2}$ and GaH$_{2}$ complexes
in Si. The origin of the vertical axis corresponds to the state $\textrm{H}_{2}+X^{-}+\textrm{h}^{+}$,
where $X$ is the corresponding acceptor.}
\end{figure*}

While the reaction mechanism between H$_{2}$ and B$^{-}$ has been
addressed already, the analogous process involving H$_{2}$ and $\textrm{Ga}^{-}$
is unchartered. In Ref.~\citep{coutinho2023} we found that formation
of a metastable (BH-H$^{-}$) state precedes subsequent reactions
toward formation of BH$_{2}^{+}$. Figure~\ref{fig1}(a) generalizes
the mechanism for either $X=\textrm{B}$ and Ga, presenting two possible
routes for attaining $X$H$_{2}^{+}$ via formation of metastable
$X$H-H$^{-}$. Hole capture should be involved in both routes. The
release of protons (not shown) with corresponding formation of additional
$X\textrm{H}$ pairs is also possible.

\noindent 

In step 1 ($\textrm{H}_{2}+X^{-}\rightarrow X\textrm{H}\textrm{-}\textrm{H}^{-}$),
the molecule strikes the $X$-Si bond and that has an energy barrier
of 1.10~eV and 1.05~eV for $X=\textrm{B}$ and Ga, respectively.
The $X\textrm{H}\textrm{-}\textrm{H}^{-}$ landing state is 0.59 and
0.89~eV above the $\textrm{H}_{2}+X^{-}$ initial state for $X=\textrm{B}$
and Ga, respectively. Its atomistic geometry is shown in Figure~\ref{fig1}(b)
for the case of GaH-H$^{-}$, where an interstitial H$^{-}$ hydride
sits next to a neutral GaH pair.

Activation energies for jumps of H$^{-}$ in both $\textrm{BH}\textrm{-}\textrm{H}^{-}$
and $\textrm{GaH}\textrm{-}\textrm{H}^{-}$ into a neighboring interstitial
cage was estimated about 0.5~eV. If we add this figure to the energy
of the $X\textrm{H}\textrm{-}\textrm{H}^{-}$ state we find 1.1~eV
and 1.4~eV for the barrier that has to be surmounted to reach the
$X\textrm{H}_{2}^{-}(\textrm{A})$ acceptor state (steps 1 and 2a
combined) for $X=\textrm{B}$ and Ga, respectively. The label `A'
emphasizes that the complex is an \emph{acceptor} and adopts an axial
(Si-H$_{\textrm{BC}}$~$X$-H$_{\textrm{AB}}$) geometry, where both
Si and $X$ are four-fold coordinated. The $X\textrm{H}_{2}^{-}(\textrm{A})$
state is 0.31~eV more stable and 0.14~eV less stable than $\textrm{H}_{2}+X^{-}$
for $X=\textrm{B}$ and Ga, respectively.

Figure~\ref{fig1}(c) shows a formation energy diagram where the
chemical potential of all elements, $\mu_{\textrm{ref}}$ (see Sec.~\ref{sec:methods}:
Methods) was chosen such that the origin of the energy scale corresponds
to the $\textrm{H}_{2}+X^{-}+\textrm{h}^{+}$ state. From this diagram
we clearly recognize the negative-$U$ ordering of donor and acceptor
transitions of both BH$_{2}$ and GaH$_{2}$ complexes. Figure~\ref{fig1}(c)
also shows that in p-type material, $X\textrm{H}_{2}^{-}$(A) complexes
are unstable and they are expected to either dissociate, or temporarily
convert into $X\textrm{H}_{2}^{+}$(D) upon hole capture (step 3a).
The geometry `D' of the donor state, Si-H$_{\textrm{BC}}$~$X$~H$_{\textrm{BC}}$-Si,
notably differs from `A' in that both H atoms sit approximately
at bond-center sites next to $X$. The latter is under coordinated,
\emph{i.e.} connects to two Si atoms only. Preliminary results show
that step 3a (hole capture by $X\textrm{H}_{2}^{-}(\textrm{A})$ accompanied
by reconfiguration to $X\textrm{H}_{2}^{+}(\textrm{D})$) has a small
barrier of the order of 0.1~eV.

From the above, the activation energy for $\textrm{H}_{2}+X^{-}+2\textrm{h}^{+}\rightarrow X\textrm{H}_{2}^{+}$
along steps 1-2a-3a is estimated as 1.1~eV and 1.4~eV for $X=\textrm{B}$
and Ga, respectively. The B-related intermediate states along the
reaction are also more stable than their Ga analogues. Hence, the
formation of $X\textrm{H}_{2}^{+}$ complexes along this route is
anticipated to be more favorable in B-doped than in Ga-doped Si.

Inspection of the one-electron structure of $X\textrm{H}\textrm{-}\textrm{H}^{-}$
reveals a deep fully occupied state within the gap, which could be
responsible for trapping holes (step 2b), thus converting the H$^{-}$
unit into H$^{0}$ or H$^{+}$. The barrier for such capture processes
is deemed small and it is neglected in the following analysis. From
NEB calculations we obtained minute barriers ($\lesssim0.2$~eV)
for conversion of $X\textrm{H}\textrm{-}\textrm{H}^{0}$ into $X\textrm{H}_{2}^{0}$(D).
The latter state is 0.08~eV and $\sim0$~eV above the $\textrm{H}_{2}+X^{-}+\textrm{h}^{+}$
initial state for $X=\textrm{B}$ and Ga, respectively (see Figure~\ref{fig1}(c)).
Assuming that hole capture will readily bring the complex into the
ground state $X\textrm{H}_{2}^{+}$(D), we conclude that route along
steps 1-2b-3b, is not only simpler, but its is also expected to involve
lower barriers after the critical step 1. Our assessment is that route
along steps 1-2b-3b is more likely to explain the conversion $\textrm{H}_{2}+X^{-}+\textrm{h}^{+}\rightarrow X\textrm{H}_{2}^{+}$,
for which we have found effective activation barriers of 1.10~eV
and 1.05~eV ascribed to step 1 for $X=\textrm{B}$ and Ga, respectively.

Although the dissociation of H$_{2}$ molecules assisted by B and
Ga show almost identical activation energies (perhaps it is slightly
more favorable when Ga is involved), the resulting BH$_{2}^{+}$ and
GaH$_{2}^{+}$ show very different electrical activity. This finding
was firstly reported in Ref.~\citep{fattah2023b} and it is clearly
shown by the formation energy diagram of Figure~\ref{fig1}(c). Whereas
BH$_{2}^{+}$ is a deep donor, GaH$_{2}^{+}$ is very shallow. Indeed,
the wavefunction of the donor state of GaH$_{2}^{0}$ is very diffuse
and spans the whole supercell. This is a strong indication that it
is an effective-mass-like donor. Our conclusion is that while both
BH$_{2}^{+}$ and GaH$_{2}^{+}$ are likely to form upon reaction
of H$_{2}$ with B and Ga, the latter is unlikely to lead to strong
carrier recombination activity.

We finally report on the dissociation mechanism of $X\textrm{H}_{2}^{+}$
complexes. This was partially addressed (for BH$_{2}^{+}$) in Ref.~\citep{coutinho2023}
and now we extend our analysis to GaH$_{2}^{+}$. The jump of one
of the H atoms in $X\textrm{H}_{2}^{+}$ into a neighboring Si-Si
bond center site ($X\textrm{H}_{2}^{+}\rightarrow X\textrm{H}\textrm{-}\textrm{H}^{+}$)
involves surmounting a barrier of $\sim\!0.7$~eV for both $X=\textrm{B}$
and Ga. These are lower than binding energies of $XH+\textrm{H}^{+}\rightarrow X\textrm{H}_{2}^{+}$
(Table~\ref{tab1}) plus migration barrier of H$^{+}$ (calculated
as 0.42~eV \citep{gomes2022}). From here we arrive at 0.82~eV and
0.92~eV for dissociation energies via $X\textrm{H}_{2}^{+}\rightarrow X\textrm{H}+\textrm{H}^{+}$
for $X=\textrm{B}$ and Ga, respectively. We conclude therefore, that
both BH$_{2}^{+}$ and GaH$_{2}^{+}$ are intermediate byproducts
along the reaction $\textrm{H}_{2}+2X^{-}+2\textrm{h}^{+}\rightarrow2X\textrm{H}$.
However, while BH$_{2}$ is suggested to be a non-radiative recombination
center (linked to LeTID in B-doped cells), GaH$_{2}$ is likely to
be harmless with respect to lifetime \citep{fattah2023b}.

\section{phosphorus-hydrogen pairs\label{sec:PH}}

\begin{figure}
\includegraphics[width=8.5cm]{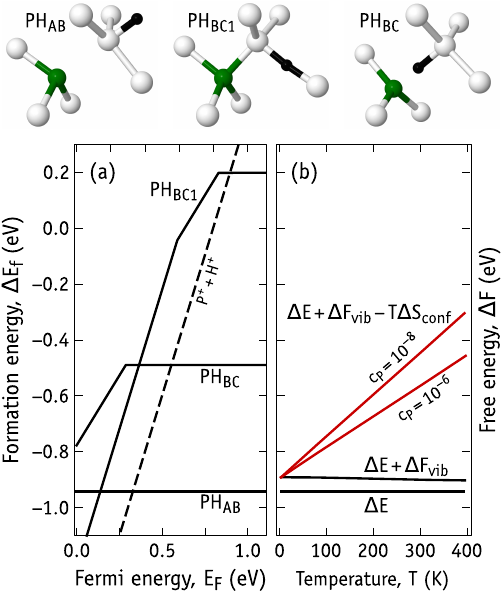}

\caption{\label{fig2}(Top) Atomistic structures of the most most stable PH
complexes in Si in the neutral and positive charge states. Phosphorus,
hydrogen and silicon are depicted in green, black and white, respectively.
(a) Formation energy of PH complexes (solid lines) as a function of
the Fermi energy. $E_{\textrm{F}}=0$ at the valence band top. The
dashed line represents the formation energy of infinitely separated
P$^{+}$ and H$^{+}$ defects. (b) Helmholtz free energy change across
the reaction $\textrm{P}^{+}+\textrm{H}^{-}\rightarrow\textrm{PH}$
in silicon (red lines). Potential energy ($\Delta E$), vibrational
Free energy ($\Delta F_{\textrm{vib}}$) and configurational entropy
($-T\Delta S_{\textrm{conf}}$) contributions are represented in a
cumulative way. Two doping concentrations are considered for the evaluation
of $S_{\textrm{conf}}$, namely $c_{\textrm{P}}=10^{-6}$ and $10^{-8}$
(see text). Both diagrams share the same vertical axis, with the origin
representing the energy of infinitely separated H$^{-}$ and P$^{+}$
impurities.}
\end{figure}

We now look at reactions between hydrogen and phosphorus. In n-type
silicon, atomic hydrogen is negatively charged and is readily attracted
by P$^{+}$ ions to form PH pairs. Among more than 10 different locations
for H next to P (up to third neighboring tetrahedral interstitial,
bond-center and anti-bonding sites), we arrived at three prominently
stable PH configurations. They are depicted in the upper part of Figure~\ref{fig2}
and are referred to as PH$_{\textrm{AB}}$, PH$_{\textrm{BC1}}$ and
PH$_{\textrm{BC}}$. Other (less stable) configurations are schematically
depicted in Figures~S1 and S2 of Supporting Information. Their respective
energies are also reported in Table~S1 of the same document. In PH$_{\textrm{AB}}$
($\equiv\textrm{P}\!:~\textrm{Si-H}$), H connects to Si along the
anti-bonding direction leaving a three-fold coordinated phosphorous
with a fully occupied lone-pair of electrons resonant with the valence
band. In PH$_{\textrm{BC1}}$ ($\equiv$P-Si-H-Si$\equiv$), H sits
at the closest Si-Si bond-center site next to the P atom. The latter
is four-fold coordinated. In PH$_{\textrm{BC}}$ ($\equiv\textrm{P}\!:~\textrm{H-Si}\equiv$)
the H atom connects to Si, next to the electronic lone pair of P.
Both PH$_{\textrm{AB}}$ and PH$_{\textrm{BC}}$ were extensively
explored in the past, the former being generally accepted as the ground
state \citep{johnson1986,chang1988,estreicher1989,chang1989,denteneer1990,estreicher1991,estreicher1994}.

Among the above geometries, only the ground state PH$_{\textrm{AB}}$
displayed a clean band gap. Calculation of electronic transitions
from total energies confirmed that PH$_{\textrm{AB}}$ is electrically
inert, chemically passivating the P dopant. The PH$_{\textrm{BC1}}$
defect has a high-lying and fully occupied level, responsible for
$(0/+)$ and $(+/+\!+)$ transitions at $E_{\textrm{c}}-0.27$~eV
and $E_{\textrm{c}}-0.51$~eV, mostly localized on the Si-H$_{\textrm{BC1}}$-Si
unit. The PH$_{\textrm{BC}}$ geometry also has a fully occupied level
in the gap, but closer to the valence band top. The origin of the
level stems from the overlap of the 1s electron of H with the P lone
pair, raising the energy of the latter above the valence band top.
From total energies we found a $(0/+)$ transition of PH$_{\textrm{BC}}$
at $E_{\textrm{v}}+0.29$~eV (no second donor transition was found).

\noindent 

Figure~\ref{fig2}(a) shows a formation energy diagram for the PH
pair in Si, where the relative energies of the three most stable configurations
are graphically combined as a function of the Fermi energy (solid
lines). The origin of the formation energy axis was chosen to be the
$\textrm{P}^{+}+\textrm{H}^{-}$ state (uncorrelated P$^{+}$ and
H$^{-}$ ions), so that binding energies are readily obtained. The
diagram shows that PH pairs adopt a neutral PH$_{\textrm{AB}}^{0}$
ground state for a wide range of Fermi level positions, most notably
for $E_{\textrm{F}}>0.33$~eV. Importantly, the results cannot explain
the observed interaction of PH pairs with minority carriers in n-type
Si \citep{seager1990}. The works of Refs.~\citep{estreicher1991,estreicher1994}
suggested that the observed charge state changes of the pairs upon
hole capture could be explained by conversion between stable PH$_{\textrm{AB}}^{0}$
and PH$_{\textrm{BC}}^{+}$ states. A requirement for this to be credible
is that PH$_{\textrm{BC}}^{+}$ would have be the ground state under
illumination or current injection conditions, \emph{i.e.}, there had
to be a $(0/+)$ transition somewhere in the gap involving PH$_{\textrm{AB}}^{0}$
and PH$_{\textrm{BC}}^{+}$. Nearly three decades ago there were no
accurate methods to find that. However, current hybrid functionals,
which avoid well known limitations of semi-local exchange-correlation
treatments, allow us to evaluate transition levels with 0.1~eV-accuracy.
Our findings, depicted in Figure~\ref{fig2}(a), clearly do not support
the proposed PH$_{\textrm{AB}}^{0}$$\longleftrightarrow$PH$_{\textrm{BC}}^{+}$
transformative model. The figure also shows that double positive charge
states are not stable in n-type Si. That includes uncorrelated $\textrm{P}^{+}+\textrm{H}^{+}$
pairs (dashed line), which are only expected to form in p-type material.

Another pillar of the transformative model is the assignment of the
$\textrm{PH}{}_{\textrm{BC}}^{0}\rightarrow\textrm{PH}{}_{\textrm{AB}}^{0}$
transformation barrier to the observed 0.7~eV activation energy of
the passivation recovery kinetics of samples in darkness (which were
previously illuminated) \citep{johnson1992}. The interpretation was
that under open-circuit and dark conditions, PH$_{\textrm{BC}}^{+}$
complexes capture electrons, allowing them to return to their $\textrm{PH}{}_{\textrm{AB}}^{0}$
ground states. Our NEB calculations for this process give a barrier
of 1.24~eV, a figure which is way too large to explain the recovery
of resistivity at room temperature.

Another puzzle relates to the calculated binding energy $E_{\textrm{b}}=0.94$~eV
for $\textrm{P}^{+}+\textrm{H}^{-}\rightarrow\textrm{PH}$, which
is larger than that of BH ($E_{\textrm{b}}=0.76$~eV), in apparent
contradiction with the higher thermal stability of the latter ---
whereas PH anneals out at about $\sim\!100$~ºC (in the dark) \citep{bergman1988},
the BH pairs dissociate in the range 140-200~ºC \citep{zundel1989}.

Figure~\ref{fig2}(b) represents the calculated Helmholtz free energy
change $\Delta F=\Delta E+\Delta F_{\textrm{vib}}-T\Delta S_{\textrm{conf}}$
across $\textrm{P}^{+}+\textrm{H}^{-}\rightarrow\textrm{PH}$ as a
function of temperature. Recent studies indicate that in n-type Si
(doping level $10^{15}\:\textrm{cm}^{-3}$), $\textrm{H}^{+}$ becomes
the dominant species in thermal equilibrium at $T\gtrsim400\:\textrm{K}$
\citep{sun2015,sun2021}. This is also an approximate upper limit
for the thermal stability of PH. The graph is therefore limited to
that temperature. For the evaluation of the free energy we considered
changes in the all-electron potential energy ($\Delta E$), vibrational
free energy ($\Delta F_{\textrm{vib}}$) and configurational entropy
($\Delta S_{\textrm{conf}}$). The latter dominates the temperature
dependence, accounting for $-T\Delta S_{\textrm{conf}}\approx0.3\textrm{-}0.4$~eV
at room temperature. The configurational entropy change (per H atom)
is approximated to $\Delta S_{\textrm{conf}}=k_{\textrm{B}}\ln(2c_{\textrm{P}})$
\citep{coutinho2023}, where $k_{\textrm{B}}$ is the Boltzmann constant,
and $c_{\textrm{P}}=n_{\textrm{P}}/n_{\textrm{Si}}$ is the fractional
concentration of phosphorus donors, $n_{\textrm{P}}$ and $n_{\textrm{Si}}$
being absolute concentrations of P and Si atoms in the crystal. Figure~\ref{fig2}(b)
considers doping concentrations of $c_{\textrm{P}}=10^{-8}$ ($n_{\textrm{P}}=5\times10^{14}$~cm$^{-3}$)
and $c_{\textrm{P}}=10^{-6}$ ($n_{\textrm{P}}=5\times10^{16}$~cm$^{-3}$).
We note that for the evaluation of $\Delta S_{\textrm{conf}}$ we
assume that the pairs are either all dissociated in the reactants
side (high temperatures) or they are connected as PH$_{\textrm{AB}}$
in the products side (low temperatures). It is also assumed that $n_{\textrm{H}}\ll n_{\textrm{P}}$
and there are no defect-defect interactions except for PH pair formation.
The results are therefore qualitative. Although anharmonic effects
are only expected to become relevant at $T\apprge500$~K \citep{estreicher2004},
we will argue that the lack of electron-phonon coupling is another
limitation in the calculations.

According to Figure~\ref{fig2}(b), at the observed annealing temperature
of PH ($T\sim350\textrm{-}400\:\textrm{K}$), the calculations indicate
that the pair is still $\sim0.4\:\textrm{eV}$ more stable than the
$\textrm{P}^{+}+\textrm{H}^{-}$ state. This is owed to the large
binding energy of the PH pair. The result contrasts with analogous
calculations for BH, where the annealing temperature was estimated
at $T\approx450$~K \citep{coutinho2023}, in fair agreement with
experiments.

\noindent 
\begin{figure*}
\includegraphics[width=18cm]{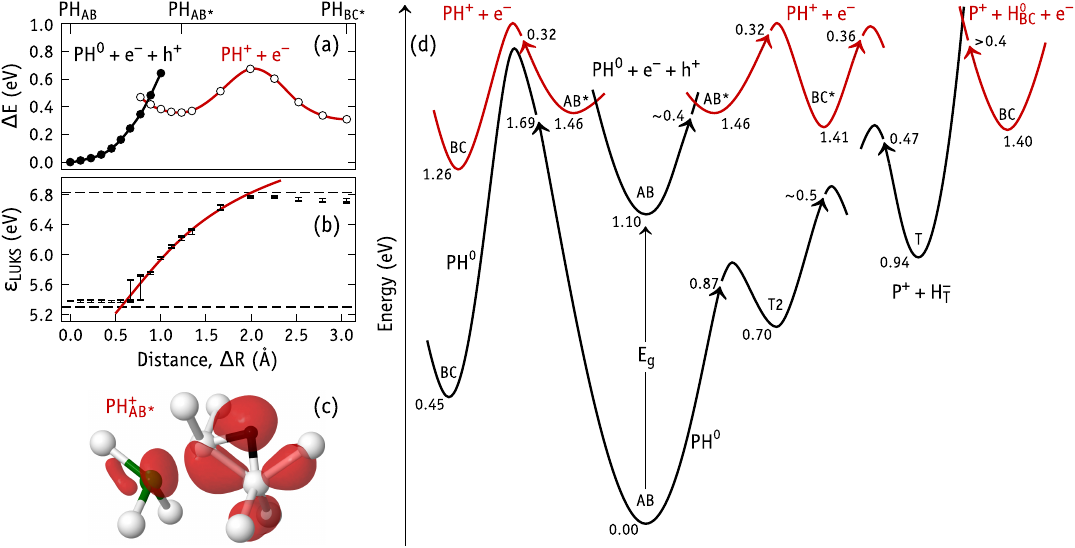}

\caption{\label{fig3}Proposed carrier-induced dissociation mechanism of PH
pairs in silicon. (a) NEB calculation of the hole capture barrier
for $\textrm{PH}_{\textrm{AB}}^{0}+\textrm{h}^{+}\rightarrow\textrm{PH}_{\textrm{BC*}}^{+}$.
Total energies of neutral and positively charged PH pairs are shown
as filled and open circles. (b) Energy of the lowest unoccupied Kohn-Sham
(LUKS) state of PH$^{+}$. Defect geometries in (a) and (b) vary from
those of $\textrm{PH}_{\textrm{AB}}^{0}$ ($\Delta R=0$~Å) to $\textrm{PH}_{\textrm{BC*}}^{+}$
($\Delta R=3.1$~Å) via $\textrm{PH}_{\textrm{AB*}}^{+}$ (see text).
The red line emphasizes the emergence of a hole trap about $\Delta R\sim0.7$~Å,
and it is drawn for the sake of eye guidance only. Dashed lines mark
the HOKS and LUKS states in a bulk supercell. (c) Isosurface (red)
of the hole localization in metastable $\textrm{PH}_{\textrm{AB*}}^{+}$
state ($\Delta R\sim1.2$~Å) calculated from the density of the LUKS
state. P, H and Si atoms are depicted in green, black and white, respectively.
(d) Configuration coordinate diagram describing interactions of PH
pairs with carriers in n-type Si. Potential energy of neutral and
positively charged pairs are shown as black and red lines, respectively.
Defect geometries and respective energies are indicated next to the
minima. Their lateral location in the diagram is not tied to a coordinate
axis. The energy minima are calculated with respect to the $\textrm{PH}_{\textrm{AB}}^{0}$
ground state. The $\textrm{PH}_{\textrm{AB}}^{0}+\textrm{e}^{-}+\textrm{h}^{+}$
state at $E_{\textrm{g}}=1.10$~eV, represents a neutral PH defect
plus an uncorrelated electron-hole pair. Energy barriers are represented
next to the arrow heads (which indicate the direction of the reaction)
and they are calculated with respect to the closest minima.}
\end{figure*}

To address the above issues we investigated the potential energy surface
of the H atom in the neighborhood of P. From the perspective of a
charge state transition, the PH$_{\textrm{BC}}^{+}$ geometry is rather
distant from PH$_{\textrm{AB}}^{0}$. Hence, we were particularly
thorough in searching for a hole trapping mechanism involving structures
closer to the PH$_{\textrm{AB}}^{0}$ ground state. We found that
small wagging oscillations of the Si-H$_{\textrm{AB}}$ unit of PH$_{\textrm{AB}}^{0}$
led to the appearance of an occupied gap level above the valence band
edge. The oscillations are depicted schematically in Figure~S1 of
Supporting Information (for the H atom sitting on site 1). That particular
movement and respective changes to the electronic structure were investigated
using the NEB method.

Figure~\ref{fig3}(a) shows potential energy diagrams for PH$^{0}$
and PH$^{+}$ with their Si-H units performing a wagging movement.
Figure~\ref{fig3}(b) represents the corresponding position of the
lowest unoccupied Kohn-Sham (LUKS) state of PH$^{+}$ within the band
gap as a function of the defect coordinate. Horizontal dashed lines
at 5.31~eV and 6.83~eV of Figure~\ref{fig3}(b) mark the energy
of the highest occupied Kohn-Sham (HOKS) and LUKS states of a 216-atom
bulk supercell at the Monkhorst-Pack special $\mathbf{k}$-point (folded
$2\times2\times2$ grid). Note that these are not the band edges that
define the indirect band gap as obtained from a primitive cell. Both
diagrams of Figures~\ref{fig3}(a) and \ref{fig3}(b) share the same
horizontal axis, which represents the cumulative distance traveled
by all atoms (mostly by H). In the present context, the LUKS of PH$^{+}$
stands for the state where a hole has been trapped. In Figure~\ref{fig3}(b)
we can notice up to three closely spaced horizontal bars for each
geometry. They refer to the LUKS states for all symmetry-irreducible
$\mathbf{k}$-points among the $2\times2\times2$ grid (see Methods
section). Their spacing gives us an idea of the dispersion of the
level in the BZ.

The itinerary of the wagging motion is $\textrm{AB}\rightarrow\textrm{AB}^{*}\rightarrow\textrm{BC}^{*}$
(and back). The $\textrm{AB}^{*}$ and $\textrm{BC}^{*}$ geometries
($\equiv\textrm{P:}\;\textrm{Si-}\textrm{H}_{\textrm{AB}}^{*}$ and
$\equiv\textrm{P-Si-}\textrm{H}_{\textrm{BC}}^{*}\textrm{-Si}\equiv$),
are metastable positive charge states, consisting of increasingly
bent Si-H and Si-H-Si units next to a three-fold and four-fold coordinated
P atoms, respectively. The PH$_{\textrm{BC*}}^{+}$ state is already
close, but still separated from PH$_{\textrm{BC1}}^{+}$ by a small
barrier (0.36~eV high). The less distorted $\textrm{PH}_{\textrm{AB*}}^{+}$
metastable structure is very close to that of $\textrm{PH}_{\textrm{AB}}^{0}$,
and the localization of the trapped hole is represented in Figure~\ref{fig3}(c).

Figure~\ref{fig3}(a) depicts the only route that we found for the
trapping of holes by $\textrm{PH}_{\textrm{AB}}^{0}$. It involves
overcoming a capture barrier of $\sim0.4$~eV before attaining $\textrm{PH}_{\textrm{AB*}}^{+}$
($R\approx1.2$~Å). Figure~\ref{fig3}(b) shows that the hole trap
mixes with the valence band top, and emerges at $R\gtrsim0.7$~Å.
Hence, a large cross-section for holes is expected (not for electrons).
After arriving at $\textrm{PH}_{\textrm{AB*}}^{+}$ (Figure~\ref{fig3}(c)),
several competing processes can follow, including hole re-emission,
$\textrm{PH}_{\textrm{AB*}}^{+}\rightarrow\textrm{PH}_{\textrm{AB}}^{0}+\textrm{h}^{+}$,
or H jumps into PH$_{\textrm{BC*}}^{+}$, PH$_{\textrm{BC1}}^{+}$
or even PH$_{\textrm{BC}}^{+}$. In the first case, the estimated
emission barrier is $\sim\!0.1$~eV and the defect returns to the
neutral ground state. Regarding the H jumps, the barriers were calculated
in the range 0.32-0.36~eV. These are rather low barriers, which should
be easily surmounted at room temperature, and ultimately lead to $\textrm{PH}^{+}\rightarrow\textrm{P}^{+}+\textrm{H}^{0}$
dissociation.

So far, none of the $\textrm{PH}{}^{+}$ states presented (with BC$^{*}$,
BC1 or BC geometries), are more stable than $\textrm{PH}_{\textrm{AB}}^{0}+\textrm{h}^{+}$,
so we still do not have an explanation for the photo- or current-induced
dissociation of the PH pairs. For that, there must be a state, which
is more stable than PH$_{\textrm{AB}}^{0}$ under illumination, and
less stable in the dark. We propose that finite temperature effects
could explain the existence of such state. From the all-electron energy
calculations, we find that the photo-/injection-induced dissociation
$\textrm{PH}{}_{\textrm{AB}}^{0}+\textrm{h}^{+}\rightarrow\textrm{P}^{+}+\textrm{H}_{\textrm{BC}}^{0}$
has a potential energy cost of $\Delta E_{\textrm{R}}=0.30$~eV.
If we account for the configurational entropy raise of that reaction,
we find that already near room temperature, $-T\Delta S_{\textrm{conf}}\approx-(0.3\textrm{-}0.4)$~eV,
meaning that the free energy of dispersed $\textrm{P}^{+}+\textrm{H}^{0}$
becomes lower than that of the pairs, hence providing favorable conditions
for the dissociation. On top of that, under persistent illumination,
a second hole capture via $\textrm{PH}_{\textrm{BC1}}^{+}+\textrm{h}^{+}\rightarrow\textrm{PH}_{\textrm{BC1}}^{+\!+}$
or via $\textrm{P}^{+}+\textrm{H}^{0}+\textrm{h}^{+}\rightarrow\textrm{P}^{+}+\textrm{H}^{+}$
(dashed line in Figure~\ref{fig2}(a)), would lead to a further drop
in the energy and to Coulomb repulsion between $\textrm{P}^{+}$ and
$\textrm{H}^{+}$, further enhancing of the dissociation rate.

Figure~\ref{fig3}(d) shows a configuration coordinate diagram that
summarizes our results for the PH pair in n-type Si. These diagrams
are usually accompanied by a horizontal axis referring to a generalized
coordinate. It would be meaningless, if not erroneous, to locate so
many different geometries against a single axis. Hence, for the sake
of diagram sanity, the coordinate axis is not shown, although the
identification of each geometry is included next to each potential
energy minimum.

Ground state potentials of close pairs (PH$^{0}$) and uncorrelated
P$^{+}$ and H$^{-}$ ions ($\textrm{P}^{+}+\textrm{H}_{\textrm{T}}^{-}$)
are shown at the bottom and right-hand side of the diagram, respectively.
At the top we find an excited state corresponding to the generation
of a free electron-hole pair ($\textrm{PH}^{0}+\textrm{e}^{-}+\textrm{h}^{+}$)
and states that result from thermally assisted capture of a minority
carrier ($\textrm{PH}^{+}+\textrm{e}^{-}$). Relative energies with
respect to ground state $\textrm{PH}_{\textrm{AB}}^{0}$ are indicated
next to the potential minima. Also indicated are several potential
energy barriers for processes of interest (next to the arrow heads).
These energies are relative to the nearest minimum. Finite temperature
effects (not represented), most notably from configurational entropy,
stabilize the states on the right hand side of the diagram, namely
$\textrm{P}^{+}+\textrm{H}_{\textrm{BC}}^{0}+\textrm{e}^{-}$ and
$\textrm{P}^{+}+\textrm{H}_{\textrm{T}}^{-}$.

At room temperature and darkness, the ground state is the $\textrm{PH}_{\textrm{AB}}^{0}$
passivated pair. Conversion to metastable $\textrm{PH}_{\textrm{BC}}^{0}$
involves surmounting a $\sim\!1.7$~eV barrier, which is even higher
than the dissociation barrier (1.41~eV). However, when minority carriers
(holes) are present and some heat is provided, a thermally-assisted
capture $\textrm{PH}_{\textrm{AB}}^{0}+\textrm{h}^{+}\rightarrow\textrm{PH}_{\textrm{AB*}}^{+}$
may occur with a capture barrier of about 0.4~eV. This provides the
opportunity of H$^{0}$ to quickly jump away from P$^{+}$, and find
a more stable state with larger entropy.

The diagram also shows that from $\textrm{PH}_{\textrm{AB*}}^{+}$,
the defect can be converted to $\textrm{PH}_{\textrm{BC}}^{+}$ by
overcoming a barrier of only 0.32~eV (toward the left hand side of
the excited state potential). From here, it could capture an electron
and become trapped at the relatively deep potential of $\textrm{PH}_{\textrm{BC}}^{0}$.
However, this picture cannot explain the increase of fixed charges
observed during light-soaked annealing treatments of hydrogenated
diodes (reactivation of $\textrm{P}^{+}+\textrm{e}^{-}$).

Our calculations support a dissociative model for the observed light-/current-induced
changes of PH pairs in Si. After dissociation of PH, and if darkness
is restored, any metastable H$_{\textrm{BC}}^{0}$ will relax to H$_{\textrm{T}}^{-}$
after electron capture. This step must be preceded by a reconfiguration
of H from the bond center site into a tetrahedral cage of the crystal,
surmounting a barrier of about 0.4~eV \citep{gomes2022}. That is
our estimated lower bound for the capture barrier of $\textrm{H}_{\textrm{BC}}^{0}+\textrm{e}^{-}\rightarrow\textrm{H}_{\textrm{T}}^{-}$
(right hand side of the diagram). The recovery of the pairs then involves
the migration of isolated H$^{-}$ toward the P$^{+}$ ions (with
a calculated 0.47~eV jumping barrier). This result is in line with
the findings of Johnson and Herring \citep{johnson1992}, which found
an activation energy of 0.7~eV for the whole recovery process. We
note that the measured barrier can be related (but not exclusively)
to an activation energy for the release of H$^{-}$ from a trapping
site (\emph{e.g.} interstitial oxygen), and as emphasized in Ref.~\citep{johnson1992},
the measured figure represents an upper bound for the migration barrier
of H$^{-}$.

The thermally-assisted capture mechanism $\textrm{PH}_{\textrm{AB}}^{0}+\textrm{h}^{+}\rightarrow\textrm{PH}_{\textrm{AB*}}^{+}$
leads us to postulate that the thermal stability of the PH pairs,
which is considerably lower than that of BH pairs, could be limited
by interactions with intrinsic carriers. At 100~ºC the intrinsic
hole concentration is already $1.5\times10^{12}$~cm$^{-3}$ \citep{sproul1991},
and their capture could accelerate PH dissociation by (1) providing
a low-barrier escape route for H (the neutral state is expected to
travel much faster than H$^{-}$), (2) turning off the Coulomb attraction
between hydrogen and phosphorus, and (3) attaining a higher entropy
state in the presence of a steady-state hole population. This mechanism
differs markedly from the dissociation of the BH pair. In that case,
close B$^{-}$ and H$^{+}$ units are electrically inactive, even
when separated by a few Si-Si bonds \citep{coutinho2023}, implying
that any light/injection-induced enhancement of BH dissociation occurs
only after enough separation of the pairs is verified, which requires
considerable heat to be provided ($T\gtrsim180~^{\circ}\textrm{C}$)
\citep{seager1991}.

We finally note that a more rigorous account of the above picture
would include the calculation of the capture cross section for $\textrm{PH}_{\textrm{AB}}^{0}+\textrm{h}^{+}\rightarrow\textrm{PH}_{\textrm{AB*}}^{+}$.
That involves finding the hole capture rate and the respective electron-phonon
matrix elements \citep{shi2012,alkauskas2014,kim2019}. This is outside
the scope of the present work. Importantly, our results suggest that
the PH pair, although it is electrically inert from the perspective
of a static calculation, that may not be the case if we account for
electron-phonon coupling.

\section{Interactions of hydrogen molecules with phosphorus\label{sec:PH2}}

From several geometries (up to 10 pair combinations of H sites as
depicted in Figs.~S1 and S2 of Supporting Information) and charge
states ($q=\{-1,\:0,\:+1\}$) of complexes that result from reaction
$\textrm{P}^{+}+\textrm{H}_{2}+(1-q)\textrm{e}^{-}\rightarrow\textrm{PH}_{2}^{q}$
between P and H$_{2}$, we could only find two stable configurations
with $\Delta E_{\textrm{R}}\lesssim+1\:\textrm{eV}$. Their geometry
consists of H pairs at sites 3-5 and 1-3 shown in Fig.~S1 of Supporting
Information. They are analogous to $X\textrm{H}_{2}^{-}(\textrm{A})$
complexes ($X$ is a group-III acceptor), but now a Si-P bond is either
replaced by $\textrm{Si-H}_{\textrm{BC}}\;\;\textrm{P-H}_{\textrm{AB}}$
or by $\textrm{H}_{\textrm{AB}}\textrm{-Si}\;\;\textrm{H}_{\textrm{BC}}\textrm{-P}$.
Both structures display trigonal ($C_{3v}$) point group symmetry,
they show close stability (the former is more stable by 0.2~eV only),
and because P is four-fold coordinated (Si and H have their \emph{normal}
coordination), both PH$_{2}$ defects are shallow donors. We can therefore
already presume, that unlike BH$_{2}$ complexes which were suggested
to be responsible for LeTID in B-doped Si, formation of PH$_{2}$
complexes is unlikely to result in strong recombination activity.

Although we did not investigate the dissociation mechanism of H$_{2}$
molecules next to P$^{+}$ dopants (as we did for B$^{-}$ and Ga$^{-}$
acceptors), we could find that such reactions are not very favorable
and some are actually endothermic. PH$_{2}$ complexes are expected
to be thermally ionized at room temperature and above, and hence,
for the sake of reaction analysis, they are considered in the positive
charge state. The results are summarized in Table~\ref{tab2}. The
first three rows account for reactions involving the formation of
PH$_{2}^{+}$ (in the most stable $\textrm{Si-H}_{\textrm{BC}}\;\;\textrm{P-H}_{\textrm{AB}}$
form). The reaction between H$_{2}$ and P$^{+}$ in the first row
is endothermic, $\Delta E_{\textrm{R}}=+0.06$~eV, and that still
excludes the fact that $\Delta S_{\textrm{R}}<0$, which makes the
reactants even more stable at finite temperatures. The reaction in
the second row shows that free $\textrm{H}^{-}$ can by trapped by
PH pairs (with a binding energy of 0.40~eV). However, this can only
occur transiently, and when the concentration of PH pairs is much
higher than that of $\textrm{P}^{+}$ ions. Otherwise, as shown by
the result of the third row, each $\textrm{PH}_{2}$ complex will
eventually dissociate to form PH pairs (with 0.47~eV gain per $\textrm{PH}_{2}$
complex).

\noindent 
\begin{table}
\caption{\label{tab2}Calculated reaction energies ($\Delta E_{\textrm{R}}$)
for formation of phosphorus-hydrogen complexes in n-type silicon.
All values are in eV.}

\begin{tabular}{lllr@{\extracolsep{0pt}.}l}
\hline 
Reaction &  &  & \multicolumn{2}{c}{$\Delta E_{\textrm{R}}$}\tabularnewline
\hline 
$\textrm{P}^{+}+\textrm{H}_{2}\rightarrow\textrm{PH}_{2}^{+}$ &  &  & \multicolumn{2}{c}{$+0.06$}\tabularnewline
$\textrm{PH}_{\textrm{AB}}^{0}+\textrm{H}_{\textrm{T}}^{-}\rightarrow\textrm{PH}_{2}^{+}+2\textrm{e}^{-}$ &  &  & \multicolumn{2}{c}{$-0.40$}\tabularnewline
$2\textrm{PH}_{\textrm{AB}}^{0}\rightarrow\textrm{PH}_{2}^{+}+\textrm{P}^{+}+2\textrm{e}^{-}$ &  &  & \multicolumn{2}{c}{$+0.47$}\tabularnewline
$2\textrm{P}^{+}+\textrm{H}_{2}+2\textrm{e}^{-}\rightarrow\textrm{P}^{+}+\textrm{PH}_{\textrm{AB}}^{0}+\textrm{H}_{\textrm{T}}^{-}$ &  &  & \multicolumn{2}{c}{$+0.46$}\tabularnewline
$2\textrm{P}^{+}+\textrm{H}_{2}+2\textrm{e}^{-}\rightarrow2\textrm{PH}_{\textrm{AB}}^{0}$ &  &  & \multicolumn{2}{c}{$-0.41$}\tabularnewline
\hline 
\end{tabular}
\end{table}

Another important question is: what does theory anticipate for the
energetics of PH pair formation from interactions between H$_{2}$
and P$^{+}$? The answer is partially found in the last two rows of
Table~\ref{tab2}. The first indicates that there is a barrier for
the full conversion, which is greater than $\sim\!1$~eV (\emph{i.e.}
0.46~eV plus the migration barrier of $\textrm{H}^{-}$). The last
row shows that a full reaction between H$_{2}$ and P$^{+}$ donors,

\begin{equation}
2\textrm{P}^{+}+\textrm{H}_{2}+2\textrm{e}^{-}\rightarrow2\textrm{PH},\label{eq:reaction1}
\end{equation}
is exothermic, although it does not lead to a substantial potential
energy drop ($\Delta E_{\textrm{R}}=-0.41$~eV). Of course, at finite
temperatures, and most importantly in the solar context, at room temperature
and above, several factors are expected to play agains the above reaction.
These include changes in configurational entropy, electronic free
energy (due to subtraction of two free-electrons), rotational and
vibrational free energies (consumption of the H$_{2}$ molecules),
as well as the capture of photogenerated holes by PH.

Let us first estimate the configurational entropy change. For Reaction~\ref{eq:reaction1}
this quantity can be approximated to $\Delta S_{\textrm{conf}}=k_{\textrm{B}}\left[\ln\left(8c_{\textrm{P}}^{2}/c_{\textrm{H}}\right)+1\right]$
(see Appendix B of Ref.~\citep{coutinho2023}), and for instance,
choosing fractional concentrations $c_{\textrm{P}}=10^{-7}$ and $c_{\textrm{H}}=10^{-10}$
($n_{\textrm{P}}=5\times10^{15}$~cm$^{-3}$ and $n_{\textrm{H}}=5\times10^{12}$~cm$^{-3}$),
we find $(-300\,\textrm{K})\times\Delta S_{\textrm{conf}}=0.16$~eV
and $(-400\,\textrm{K})\times\Delta S_{\textrm{conf}}=0.21$~eV.

The raise in the Helmholtz electronic free energy across the reaction
is estimated from $\Delta F_{\textrm{elec}}\approx f(\mu-k_{\textrm{B}}T)$,
which subtracts $pV=k_{\textrm{B}}T$ (for an ideal electron gas)
to the Gibbs free energy per free-electron $\mu\approx k_{\textrm{B}}T\ln(\Delta n/N_{\textrm{c}})$
(see Ref.~\citep{estreicher2004}), where $\Delta n=fn_{\textrm{H}}$
is the effective change in the free-electron density, $f=\exp(-E_{\textrm{i}}/k_{\textrm{B}}T)$
is a Boltzmann factor quantifying the effective fraction of thermally
ionized phosphorus donors with ionization energy $E_{\textrm{i}}$,
and $N_{\textrm{c}}$ is the effective density of states at the bottom
of the conduction band. From the above, we find $\Delta F_{\textrm{elec}}=0.12$~eV
and 0.24~eV at $T=300$~K and 400~K, respectively, meaning that
even without considering the roto-vibrational contribution of the
molecule, the free energy change $\Delta F=\Delta E_{\textrm{R}}+\Delta F_{\textrm{elec}}-T\Delta S_{\textrm{conf}}$
across Reaction~\ref{eq:reaction1} is nearly zero at room temperature.

Our results suggest that after cooling P-doped Si that was in contact
with a high-temperature hydrogen source, \emph{e.g.} after a fast-firing
step for contact formation during solar cell fabrication, PH pair
formation is unlikely to occur at the expense of direct reactions
between H$_{2}$ molecules and P$^{+}$. This conclusion is indirectly
supported by the measurements of Pritchard \emph{et~al.} \citep{pritchard1998},
who followed the detachment of H$_{2}$ molecules from interstitial
oxygen defects (which act as trapping sites) in P-doped Czochralski
Si samples, showing a concomitant increase of free H$_{2}$ located
at tetrahedral interstitial sites of the crystal. The conversion was
monitored during annealing treatments between room temperature and
$T=130$~ºC, there was no indication of PH pair formation in the
initial state (\emph{as-quenched}) of the samples, and the conversion
from trapped to free H$_{2}$ was complete, irrespectively of the
temperature. This contrasts with analogous experiments in B-doped
material, where a substantial fraction of H was already present in
the form of BH pairs after quenching the samples (previously put in
contact with an H$_{2}$ gas at 1200~ºC) to room temperature \citep{pritchard1999}.

\section{Hydrogen reactions with carbon\label{sec:CH}}

Here we have a look into several open issues related to carbon-hydrogen
complexes in Si, especially regarding their formation, dissociation,
and their electronic activity. Interactions between hydrogen and substitutional
carbon is likely to occur in solar-grade Si, where the concentration
of substitutional carbon can reach $10^{16}$~cm$^{-3}$. Complexes
that result from hydrogenation of substitutional carbon pairs ($\textrm{C}_{2}\textrm{H}_{n}$)
are only expected in C-rich material, and they are left outside the
scope of the present work.

\begin{figure}
\includegraphics[width=8.5cm]{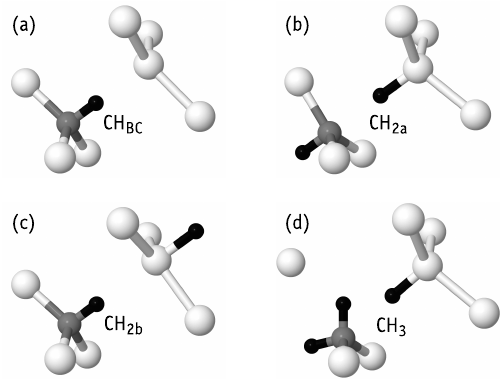}

\caption{\label{fig4}Atomistic structures of the most most stable CH$_{n}$
complexes in Si ($n\protect\leq3$). Carbon, hydrogen and silicon
are depicted in gray, black and white, respectively. $\Delta E_{\textrm{R}}=E_{\textrm{fs}}-E_{\textrm{is}}$
and $\Delta E_{\textrm{A}}=E_{\textrm{ts}}-E_{\textrm{is}}$, where
subscripts \textquoteleft is\textquoteright , \textquoteleft ts\textquoteright{}
and \textquoteleft fs\textquoteright{} stand for initial, transition
and final states of the reaction on the leftmost column, respectively.
All values are in eV.}
\end{figure}

We first focus on the CH pair. According to our results the $\equiv\textrm{C-H}_{\textrm{BC}}\;\;\textrm{Si}\equiv$
configuration (referred to as $\textrm{CH}_{\textrm{BC}}$ and shown
in Figure~\ref{fig4}(a)), is the most stable for charge states $+$,
$0$ and $-$. These possess a Si dangling bond with electron occupancy
of 0, 1 and 2 electrons, respectively. The alternative structure $\textrm{H}_{\textrm{AB}}\textrm{-C}\;\;\textrm{Si}\equiv$
(referred to as $\textrm{CH}_{\textrm{AB}}$, and also showing four-
and three-fold coordinated C and Si atoms, respectively) is 0.71,
0.41 and 0.17~eV above $\textrm{CH}_{\textrm{BC}}^{+}$, $\textrm{CH}_{\textrm{BC}}^{0}$
and $\textrm{CH}_{\textrm{BC}}^{-}$ ground states, respectively.
These results are in line, but also improve upon previous local density
approximated calculations \citep{leary1998,andersen2002}. Like in
Ref.~\citep{andersen2002}, we also found low energy configurations
for $\textrm{CH}^{+}$ and $\textrm{CH}^{-}$ where H is not directly
attached to the C atom. The positive metastable state consists of
a Si-H$_{\textrm{BC}}^{+}$-Si unit next to substitutional carbon
(0.21~eV above the CH$^{+}$ ground state). This is hereafter referred
to as $\textrm{CH}_{\textrm{BC1}}^{+}$. The negative metastable state
is attained when H sits close to the tetrahedral interstitial site
next to C along the $\langle100\rangle$ direction (0.20~eV above
the CH$^{-}$ ground state). This state is analogous to that of site
10 in Figure~S1 of Supporting Information, and it is referred to
as $\textrm{CH}_{\textrm{T1}}^{-}$.

\noindent 

The binding of free hydrogen ions to carbon in p-type and n-type Si
($\textrm{H}^{+}+\textrm{C}\rightarrow\textrm{CH}^{+}$ and $\textrm{H}^{-}+\textrm{C}\rightarrow\textrm{CH}^{-}$)
is favored by $\Delta E_{\textrm{R}}=-0.41$~eV and $-0.71$~eV,
respectively. These values contrast with the not so favorable reaction
energies to produce CH upon interaction of H$_{2}$ molecules with
C (first three rows of Table~\ref{tab3}). Such difference could
explain why CH complexes form underneath the surface of wet-etched
or H-plasma treated Si, but as far as we are aware, no one has reported
their appearance after cooling hydrogenated Si from above $\sim700$~ºC
(a procedure which promotes the formation of H$_{2}$ molecules).

Interestingly, we found that $\textrm{BH}+\textrm{C}\rightarrow\textrm{B}^{-}+\textrm{CH}_{\textrm{BC}}^{+}$
and $\textrm{PH}+\textrm{C}\rightarrow\textrm{P}^{+}+\textrm{CH}_{\textrm{BC}}^{-}$
are endothermic with $\Delta E_{\textrm{R}}=+0.46$ and $+0.70$~eV,
respectively (see Table~\ref{tab3}). These figures tell us that
CH pairs cannot be formed at the expense of dopant-hydrogen pairs.

The calculated donor and acceptor transitions for $\textrm{CH}_{\textrm{BC}}$
are estimated at $E_{\textrm{v}}+0.37$~eV and $E_{\textrm{c}}-0.10$~eV,
respectively. The acceptor level agrees well with the E3 electron
trap at $E_{\textrm{c}}-0.16$~eV, measured by DLTS in n-type Si
by Kamiura \emph{et~al.} \citep{kamiura1997} (also referred to as
(C-H)$_{\textrm{II}}$ in Ref.~\citep{andersen2002} and E90 in Ref.~\citep{kolkovsky2018}),
whereas the calculated donor transition fits well the H1 hole trap
also measured by DLTS at $E_{\textrm{v}}+0.33$~eV, and assigned
to a CH defect \citep{kamiura1995} (also labeled H180 in Ref.~\citep{kolkovsky2018}).
Although the H1/H180 hole trap did no show a shift of the emission
rate with varying the bias (as it is expected for a donor in p-type
Si) \citep{kamiura1995}, the existence of a capture barrier of $\sim\!0.05$~eV
led to the suggestion that it should be an acceptor \citep{kolkovsky2018}.
Our results do not support this view.

DLTS and C-V measurements of wet-etched B-doped samples show that
H180 and BH pairs have the same depth profile \citep{stuebner2016},
thus suggesting that both also have the same number of of hydrogen
atoms. The H1 trap was shown to anneal out in the dark above 100~ºC
with an activation energy of 1.7~eV. From NEB calculations, we find
that a $\textrm{CH}_{\textrm{BC}}^{+}\rightarrow\textrm{CH}_{\textrm{BC1}}^{+}$
jump has a barrier of 1.61~eV. Assuming that subsequent $\textrm{H}{}^{+}$
jumps into farther Si-Si bonds have lower barriers, we find in this
result further support for the assignment of H1 to to the donor transition
of $\textrm{CH}_{\textrm{BC}}$.

Additional NEB calculations were performed to understand the dissociation
of $\textrm{CH}_{\textrm{BC}}$. In the neutral charge state, the
H jump with the lowest barrier (that is not a reorientation) was also
$\textrm{CH}_{\textrm{BC}}^{0}\rightarrow\textrm{CH}_{\textrm{BC1}}^{0}$,
with the saddle point at 1.41~eV above the initial state. On the
other hand, in the negative charge state the easiest first step for
dissociation was found $\textrm{CH}_{\textrm{BC}}^{-}\rightarrow\textrm{CH}_{\textrm{T1}}^{-}$,
with a much lower potential energy barrier, estimated as 0.66~eV.
These figures match very well the activation energies for annealing
of the E3 electron trap in n-type Si diodes under (1) bias and darkness
and (2) at zero-bias under white light illumination \citep{kamiura1997}.
In (1) the Fermi level was well bellow the $E_{\textrm{c}}-0.16$~eV
acceptor level and the defect did not have access to free electrons.
Under these conditions, E3 defect traps were empty (here interpreted
as the $\textrm{CH}_{\textrm{BC}}^{0}$ state) and it was shown to
take days to anneal them out at $\sim\!60$~ºC. An activation energy
of 1.33~eV (with a pre-factor of $10^{14}$~s$^{-1}$) was extracted
from isothermal annealing data assuming first-order and Arrhenius
behavior. That is very close to the calculated barrier for $\textrm{CH}_{\textrm{BC}}^{0}\rightarrow\textrm{CH}_{\textrm{BC1}}^{0}$
and the pre-factor suggests that the mechanism involves simple atomic
motion. In (2), the E3 traps had access to a non-equillibrium population
of photogenerated electrons and they annealed out in a time-scale
of minutes, even below room temperature. The annealing rate under
these conditions was estimated as $10^{6}~\textrm{s}^{-1}\times\exp(-0.5~\textrm{eV}/k_{\textrm{B}}T)$.
Now the barrier is close to that of $\textrm{CH}_{\textrm{BC}}^{-}\rightarrow\textrm{CH}_{\textrm{T1}}^{-}$,
and the pre-factor suggests that an electron is captured by $\textrm{CH}_{\textrm{BC}}^{0}$
before performing the jump.

The $\textrm{CH}_{\textrm{BC1}}$ defect has also been investigated
before in n-type Si \citep{andersen2002}. For this defect we find
a calculated donor transition at $E_{\textrm{c}}-0.22$~eV. Hence,
we support its assignment to an electron trap labeled (C-H)$_{\textrm{I}}$
(a precursor to (C-H)$_{\textrm{II}}$), measured at $E_{\textrm{c}}-0.22$~eV,
showing a clear Poole-Frenkel behavior, and ascribed to a Si-H-Si
defect next to a carbon atom \citep{andersen2002}.

As mentioned already, $\textrm{CH}_{\textrm{AB}}$ and $\textrm{CH}_{\textrm{T1}}$
geometries are quite stable in the negative charge state (0.17~eV
and 0.20~eV above $\textrm{CH}_{\textrm{BC}}^{-}$) . Like $\textrm{CH}_{\textrm{BC1}}^{+}$
in proton-implanted Si, they could be precursors to the $\textrm{CH}_{\textrm{BC}}^{-}$
ground state, especially in n-type wet-etched Si. For $\textrm{CH}_{\textrm{AB}}$
we find $(0/+)$ and $(-/0)$ transitions at $E_{\textrm{v}}+0.08$~eV
and $E_{\textrm{c}}-0.31$~eV. These are about 0.2~eV lower in the
gap than the analogous levels of $\textrm{CH}_{\textrm{BC}}$. This
shift toward lower energies can be explained by the absence of repulsion
between the 1s state of H (which is now at the anti-bonding site)
and electrons on the Si radical state. The calculated levels are also
in line with transitions arising from other Si dangling bond defects,
\emph{e.g.}, VOH with donor and acceptor transitions at $E_{\textrm{v}}+0.28$~eV
and $E_{\textrm{c}}-0.31$~eV \citep{coutinho2003}. The lowest unoccupied
electronic state of $\textrm{CH}_{\textrm{AB}}^{-}$ is a conduction
band state, and as expected, no second acceptor level was found for
this geometry.

As for $\textrm{CH}_{\textrm{T1}}$ we find a $(-/0)$ level at $E_{\textrm{v}}+0.19$~eV.
Also for this defect, the calculations suggest that it cannot trap
a second electron (it is a single acceptor). Judging from the estimated
error bar of these calculations ($\sim\!0.1$~eV), we are not able
connect any of the calculated levels of $\textrm{CH}_{\textrm{AB}}$
and $\textrm{CH}_{\textrm{T1}}$ defects to other observed traps that
were convincingly shown to be C-H related \citep{kolkovsky2018}.
Perhaps the $(-/0)$ level estimated at $E_{\textrm{c}}-0.31$~eV
for $\textrm{CH}_{\textrm{AB}}$ is not that far from the observed
deep electron trap at $E_{\textrm{c}}-0.51$~eV (labeled E262) \citep{stuebner2016b}.
Such discrepancy could in principle be explained by the existence
of an unusually large capture barrier ($\gtrsim$0.1~eV) for $\textrm{CH}_{\textrm{AB}}^{0}+\textrm{e}^{-}\rightarrow\textrm{CH}_{\textrm{AB}}^{-}$
which was not considered theoretically. However, E262 was shown to
be accompanied by another trap (labeled E46) at 0.06~eV below $E_{\textrm{c}}$,
both displaying identical depth profiles to that of the PH pair, showing
similar annealing behavior, and also showing identical dependence
on the carbon concentration \citep{stuebner2016b}. For that, E262
and E46 were assigned to first and second acceptor transitions of
the same complex, possibly $\textrm{CH}_{\textrm{AB}}$. Our calculations
seem to rule out that possibility.

\noindent 
\begin{table}
\caption{\label{tab3}Calculated reaction energies ($\Delta E_{\textrm{R}}$)
involving several carbon-, boron, and phosphorus-hydrogen complexes
in silicon. All values are in eV.}

\begin{tabular}{lllr@{\extracolsep{0pt}.}l}
\hline 
Reaction &  &  & \multicolumn{2}{c}{$\Delta E_{\textrm{R}}$}\tabularnewline
\hline 
$\frac{1}{2}\textrm{H}_{2}+\textrm{C}\rightarrow\textrm{CH}$ &  &  & \multicolumn{2}{c}{$+0.12$}\tabularnewline
$\frac{1}{2}\textrm{H}_{2}+\textrm{C}+\textrm{h}^{+}\rightarrow\textrm{CH}^{+}$ &  &  & \multicolumn{2}{c}{$-0.25$}\tabularnewline
$\frac{1}{2}\textrm{H}_{2}+\textrm{C}+\textrm{e}^{-}\rightarrow\textrm{CH}^{-}$ &  &  & \multicolumn{2}{c}{$+0.02$}\tabularnewline
$\textrm{H}_{2}+\textrm{C}\rightarrow\textrm{CH}_{2}$ &  &  & \multicolumn{2}{c}{$-0.97$}\tabularnewline
$\frac{3}{2}\textrm{H}_{2}+\textrm{C}\rightarrow\textrm{CH}_{3}$ &  &  & \multicolumn{2}{c}{$-0.90$}\tabularnewline
$\textrm{BH}+\textrm{C}\rightarrow\textrm{B}^{-}+\textrm{CH}^{+}$ &  &  & \multicolumn{2}{c}{$+0.46$}\tabularnewline
$\textrm{PH}+\textrm{C}\rightarrow\textrm{P}^{+}+\textrm{CH}^{-}$ &  &  & \multicolumn{2}{c}{$+0.70$}\tabularnewline
$2\textrm{BH}+\textrm{C}\rightarrow2\textrm{B}^{-}+\textrm{CH}_{2}+2\textrm{h}^{+}$ &  &  & \multicolumn{2}{c}{$+0.35$}\tabularnewline
$2\textrm{PH}+\textrm{C}\rightarrow2\textrm{P}^{+}+\textrm{CH}_{2}+2\textrm{e}^{-}$ &  &  & \multicolumn{2}{c}{$+0.49$}\tabularnewline
$\textrm{BH}+\textrm{CH}_{2}\rightarrow\textrm{B}^{-}+\textrm{CH}_{3}^{+}$ &  &  & \multicolumn{2}{c}{$+0.35$}\tabularnewline
$\textrm{PH}+\textrm{CH}_{2}\rightarrow\textrm{P}^{+}+\textrm{CH}_{3}^{0}+\textrm{e}^{-}$ &  &  & \multicolumn{2}{c}{$+0.80$}\tabularnewline
\hline 
\end{tabular}
\end{table}

\section{Hydrogen multi-trapping at carbon\label{sec:CH2}}

Substitutional carbon in Si is known to trap at least two hydrogen
atoms. This effect has been found both in proton implanted material
\citep{holbech1993}, and in Si samples heated above 1300~ºC in a
H$_{2}$-rich atmosphere and quenched to room temperature \citep{suezawa1999}.
As for modeling the $\textrm{CH}_{2}$ complexes, that was extensively
addressed by Estreicher and co-workers \citep{estreicher1999,estreicher2012}.
Like for the PH$_{2}$ complexes, there are two stable configurations
for $\textrm{CH}_{2}$. They are shown in Figures~\ref{fig4}(b)
and \ref{fig4}(c), and we confirm that they are nearly degenerate
and electrically inert. Their detection relies on local vibrational
mode spectroscopy only \citep{holbech1993,suezawa1999}.

Table~\ref{tab3} shows few possible reactions leading to formation
of CH$_{2}$, where we can find a substantial potential energy drop
for $\textrm{H}_{2}+\textrm{C}\rightarrow\textrm{CH}_{2}$ ($\Delta E_{\textrm{R}}=-0.97$~eV).
This reaction was studied in detail in Ref.~\citep{coutinho2023},
where it was found that like boron, carbon can enhance the dissociation
of H$_{2}$ molecules (dissociation barrier of 1.35~eV). However,
unlike B, the state attained after dissociation, Si-H H-Si, is electrically
inert and subsequent steps cannot benefit from the capture of carriers.

Like for the CH pair, formation of $\textrm{CH}_{2}$ upon release
of H from PH in n-type and BH in p-type Si is not favorable, \emph{i.e.}

\[
2\textrm{BH}+\textrm{C}\rightarrow2\textrm{B}^{-}+\textrm{CH}_{2}+2\textrm{h}^{+}
\]
and

\[
2\textrm{PH}+\textrm{C}\rightarrow2\textrm{P}^{+}+\textrm{CH}_{2}+2\textrm{e}^{-}
\]
are endothermic reactions. This suggests that CH$_{2}$ cannot be
obtained upon annealing of BH and PH pairs. Similar conclusions can
be drawn for the formation of CH$_{3}$ complexes (see Table~\ref{tab3}).
The latter complex is not even stable against decomposition into CH$_{2}$
plus a dopant-hydrogen pair. The most stable form of $\textrm{CH}_{3}$
is depicted in Figure~\ref{fig4}(d), and comprises a Si radical
next to $=\!\textrm{CH}{}_{2}$ and $\equiv\!\textrm{SiH}$ units.
A deep donor transition was calculated at $E_{\textrm{v}}+0.43$~eV
(no acceptor levels were found), and also in this case, we cannot
find a match with any of the DLTS traps summarized in Table~1 of
Ref.~\citep{kolkovsky2018}, in particular with the one labeled $\textrm{E}90'$
at $E_{\textrm{c}}-0.14$~eV assigned to a $\textrm{CH}_{n}$ defect
with $n>1$.

\section{Conclusions\label{sec:conclusions}}

We presented a comprehensive theoretical study of hydrogen-dopant
and hydrogen-carbon interactions in silicon using state-of-the-art
electronic structure methods. The impact and role of several hydrogen-related
complexes was addressed in the context of non-radiative recombination
of carriers by defects in solar silicon, in particular of LeTID of
Si cells.

The interaction of H$_{2}$ molecules with B and Ga acceptors was
investigated comparatively. We found that both $X=\{\textrm{B},\,\textrm{Ga}\}$
group-III elements act as catalysts for H$_{2}$ dissociation, leading
to formation of intermediate $X\textrm{H}_{2}^{+}$ complexes before
attaining a lower energy state consisting of acceptor-hydrogen pairs
($X$H). The activation energy of the acceptor-assisted dissociation
of H$_{2}$ is estimated 1.10~eV and 1.05~eV for B and Ga (to be
compared with 1.6~eV for H$_{2}$ dissociation in pristine Si). These
values are close to the activation energy for the LeTID development
in Si cells. These barriers are also the critical steps for formation
of BH$_{2}^{+}$ and GaH$_{2}^{+}$ along $\textrm{H}_{2}+2X^{-}+2\textrm{h}^{+}\rightarrow X\textrm{H}_{2}^{+}+X^{-}\rightarrow2X\textrm{H}$.
The BH$_{2}^{+}$ was previously assigned to the route-cause of LeTID
in B-doped solar cells. However, GaH$_{2}^{+}$ are effective-mass-like
shallow donors, and therefore, unlikely to lead to analogous recombination
activity in cells based on Ga-doped substrates.

We find that light-/carrier-induced dissociation of PH pairs cannot
be explained by a transformative model, where H jumps between anti-bonding
and bond-center sites upon capture of minority and majority carriers.
Instead, our results suggest a dissociative mechanism, triggered by
a metastable hole trap accessible to the ground state via wagging
vibrations of the Si-H$_{\textrm{AB}}$ unit. From there, the height
of the potential energy barriers for H detachment are a few tenths
of eV only, and H can either escape as H$^{0}$ or as H$^{+}$ if
it captures a another hole. In the latter case, the escape would be
enhanced by the repulsive field of P$^{+}$. Interestingly, the above
model suggests that a defect, which according to a static calculation
does not have electrical levels in the gap, is still capable of trapping
free carriers, but that can only be explained if we account for electron-phonon
coupling and finite temperature effects.

Indeed, an important contribution to the dissociation of PH is the
raise of configurational entropy. At $T=300$~K and in the presence
of minority carriers, the magnitude of $-T\Delta S_{\textrm{conf}}$
makes the reaction,

\[
\textrm{PH}^{0}+\textrm{h}^{+}\longrightarrow\textrm{P}^{+}+\textrm{H}^{0}
\]
more likely and a subsequent electron capture by hydrogen leads to
further stabilization. The above mechanism for minority carrier enhanced
dissociation of PH could explain why its annealing temperature is
$\sim\!100$~ºC lower than that of BH pairs, despite the smaller
binding energy of the latter. With respect to that, we suggest that
PH dissociates above $T\approx100$~ºC in the dark with help of the
increasing concentration of intrinsic holes.

Interactions between H$_{2}$ molecules and phosphorus was also investigated.
Direct interactions via $\textrm{H}_{2}+\textrm{P}^{+}\rightarrow\textrm{PH}_{2}^{+}$
are not favorable ($\Delta E_{\textrm{R}}=+0.06$~eV). Although the
products are the most stable form of PH$_{2}$, they are shallow donors,
and even if they could form, they are not expected to lead to recombination
activity, not even to changes in conductivity.

We found that PH pair formation at the expense of H$_{2}$ molecules
and P donors leads to a small energy drop of $\Delta E_{\textrm{R}}=-0.4$~eV,
i.e.,

\[
2\textrm{P}^{+}+\textrm{H}_{2}+2\textrm{e}^{-}\xrightarrow{\Delta E_{R}=-0.41}2\textrm{PH}.
\]
However, we also find that configurational and electronic entropy
alone (without considering roto-vibrational contributions from the
H$_{2}$ molecule, which favors the reactants side), are able to cancel
$\Delta E_{\textrm{R}}$ at $T\apprge300\textrm{-}400$~K. This result
suggests that PH formation from direct interactions between H$_{2}$
and P$^{+}$ is unlikely during the cooling of n-type solar cells
subject to fast-firing treatments.

We also explored the details of mechanisms behind the annealing of
CH pairs under dark conditions and under white light illumination
(or carrier injection). Strong differences stem from different barriers
and jump mechanisms as a function of the charge state. For CH$^{+}$,
CH$^{0}$ and CH$^{-}$, dissociation barriers were estimated as 1.61~eV,
1.41~eV and 0.66~eV, respectively. These quantities agree fairly
well with the available experimental data, shedding light into an
old and unsolved puzzle. We confirm the assignment of two measured
carrier traps (H1\citep{kamiura1995} and E3/(C-H)$_{\textrm{II}}$/E90
\citep{kamiura1997,andersen2002,kolkovsky2018}) to electronic transitions
involving the most stable configuration of the CH pair. The calculation
of a metastable donor transition involving a C-Si-H-Si structure (reminiscent
of the $(0/+)$ transition of isolated bond-centered hydrogen), also
supports its previous assignment to the (C-H)$_{\textrm{I}}$ electron
trap \citep{andersen2002}.

CH$_{n}$ complexes are not stable against formation of dopant-H pairs.
They could however form transiently in C-rich Si, \emph{e.g.}, under
solar cell operating conditions due to light-enhanced dissociation
of dopand-H complexes. The CH pair in p-type Si is positively charged,
it is stable above room temperature, and could act as a non-radiative
recombination center by attracting minority carriers (photo-generated
electrons). It is therefore a suspect to consider as LeTID defect
in cells based on B- and Ga-doped substrates. CH$_{2}$ is electrically
inert and CH$_{3}$, although has a deep donor level, it is unstable
and unlikely to form.

\section{Methods\label{sec:methods}}

We employed the density functional Vienna Ab-initio Simulation Package
(VASP) \citep{kresse1993,kresse1996a,kresse1996b}, which uses the
projector-augmented wave method \citep{bloechl1994} and planewaves
for the description of core and valence electronic states, respectively.
The maximum kinetic energy of the planewaves was 400~eV. Total energies
were evaluated self-consistently, using the hybrid density functional
of Heyd-Scuseria-Ernzerhof (HSE06) \citep{heyd2003,krukau2006}, with
a numerical accuracy of $10^{-6}$~eV. Mixing and screening parameters
were those originally proposed for this functional ($a=1/4$ and $\omega=0.2~\textrm{Å}^{-1}$)
\citep{krukau2006}, leading to an indirect band gap in the Kohn-Sham
band structure of about 1.1~eV. All-electron energies were found
for defective supercells of Si, constructed by replication of $3\times3\times3$
conventional unit cells with theoretical lattice constant $a_{0}=5.4318$~Å
(216 Si atoms). Defect structures were optimized by minimization of
the Hellmann-Feynman forces within the HSE06-level, until the largest
force became lower than 0.01~eV/Å. The band structure was sampled
on a $2\times2\times2$ Monkhorst-Pack grid of $\mathbf{k}$-points.
Such sampling level leads to well converged formation energies, with
a numerical accuracy of the order of 10~meV (see for example Ref.~\citep{shim2005}).

Formation energies ($E_{\textrm{f}}$) and transition levels of defects
(for instance with respect to the valence band top, $E(q/q')-E_{\textrm{v}}$)
were evaluated using the usual respective methodologies \citep{freysoldt2014},

\[
E_{\textrm{f}}(q,E_{\textrm{F}})=E_{\textrm{def}}(q)-\mu_{\textrm{ref}}+q(\epsilon_{\textrm{v}}+E_{\textrm{F}})
\]
and

\[
E(q/q')-E_{\textrm{v}}=[E_{\textrm{def}}(q)-E_{\textrm{def}}(q')]/(q'-q)-\epsilon_{\textrm{v}},
\]
where $E_{\textrm{def}}(q)$ is the total energy of the defect in
charge state $q$ (which already includes a periodic charge correction
\citep{freysoldt2009}), $\mu_{\textrm{ref}}$ is a reference energy
for a system with the same stoichiometry of the defective supercell,
and the term $q(\epsilon_{\textrm{v}}+E_{\textrm{F}})$ accounts for
$-q$ electrons (or $q$ holes) trapped at the defect. Finally, $\epsilon_{\textrm{v}}$
is the highest occupied state of bulk Si at $\mathbf{k}=\Gamma$ and
$E_{\textrm{F}}$ is the Fermi energy (independent variable).

For the calculation of $\mu_{\textrm{ref}}$, energies of Si, H, B,
Ga, P and C species were found from the energies per atom in bulk
Si, molecular $\textrm{H}_{2}$ in Si, substitutional B, Ga, P and
C in Si. For instance, the energy per substitutional species $X$
was found from $\mu_{X}=E(\textrm{Si}_{215}X)-215\mu_{\textrm{Si}}$,
where $E(\textrm{Si}_{215}X)$ is the energy of a 216 atom cell where
one of the Si atoms is replaced by $X$, and $\mu_{\textrm{Si}}$
is the energy per Si atom in bulk Si (silicon chemical potential).

The comparison of calculated transition levels of defects in semiconductors
with experimental data, notably with levels measured by deep level
transient spectroscopy \citep{henry1977,peaker2018}, is not always
straightforward. The calculations are usually carried out at $T=0\:\textrm{K}$
(like we did), they assume that the $T$-dependence of the energy
of both charge states is identical, and that they benefit from cancelation
effects. This condition is however not warranted, and as pointed out
by Wickramaratne \emph{et~al.} \citep{wickramaratne2018}, temperature
dependencies of the carrier capture cross section might induce a gentle
\emph{bowing} to the Arrhenius-like plot of the $T^{2}$-corrected
emission rate against $1/k_{\textrm{B}}T$. This effect was estimated
to lead to variations in the extracted activation energy of the order
of $\sim0.1$~eV across a temperature window of hundreds of Kelvin.
For defects with emission peaks around $\sim\!100$~K, such finite-temperature
effects become smaller and the measured activation energies for carrier
emission are closer to the 0~K extrapolation. The bowing is normally
unnoticed simply because the range of temperatures that are allowed
by the measurement conditions is limited.

Another effect to consider is the existence of a capture barrier.
If the experiment involves the measurement of emission rates only,
a capture barrier must be subtracted from the emission activation
energy in order to find a level position \citep{henry1977,peaker2018}.
This barrier is also temperature dependent, it is usually small ($\lesssim0.1\:\textrm{eV}$)
for defects with similar geometries in both charge states of the transition,
and decreases substantially at cryogenic temperatures (below $\sim\!100$~K)
due to tunneling effects \citep{alkauskas2014}. This effect should
always be considered with care, especially when dealing with transitions
involving a change in the defect structure. Many defects in Si have
levels in the range of up to $\sim0.4$~eV from the band edges, and
temperature effects during the measurements are not as severe as in
wide gap materials, where emission peaks are observed at few hundred
Kelvin. Of course, calculating the temperature dependent capture cross
section \citep{shi2012,alkauskas2014,kim2019}, and casting it in
the form of an Arrhenius relation \citep{wickramaratne2018}, would
bring the calculations closer to what is actually measured. However,
as referred at the end of Sec.~\ref{sec:PH}, we leave this task
for future work.

Temperature-dependent free energies of defects were also estimated
within several limitations, including the harmonic and dilute approximations,
where anharmonicity and defect-defect interactions are neglected.
In these calculations we account for the electronic potential energy,
zero-point motion, as well as vibrational, rotational and configurational
degrees of freedom. Further details have been reported elsewhere \citep{coutinho2023}.

The potential energy landscape of reactions was investigated using
the nudged elastic band (NEB) method \citep{henkelman2000}. Firstly,
up to 12 intermediate geometries, all connected by the \emph{elastic
band} between the HSE06-level end-state geometries, were relaxed.
These calculations were performed within the generalized gradient
approximation \citep{perdew1996}. Secondly, single-point calculations
at HSE06-level were performed for all geometries along the chain.
From here, we found the minimum energy path (MEP) for the reaction
of interest.

The above methodology relies on transition-state-theory, where the
reactants are assumed to convert into the products once the saddle
point -- the highest energy state along the minimum energy path between
the ends of the reaction -- is achieved, the charge and magnetic
state of the system is conserved along the way, and electron phonon
coupling is neglected. Experimental conditions and limitations also
hinder a direct comparison between measured and calculated figures.
Important effects include back-reactions, interactions with intrinsic
carriers, among other effects, which can only be acknowledged in the
analysis, or roughly accounted for.

\section*{Conflict of Interest}

The authors declare no conflict of interest.

\section*{Data Availability Statement}

The data that support the findings of this study are available at
request from the corresponding author.

\section*{Table of Contents}

\noindent 
\begin{figure}[H]
\includegraphics[width=5.5cm]{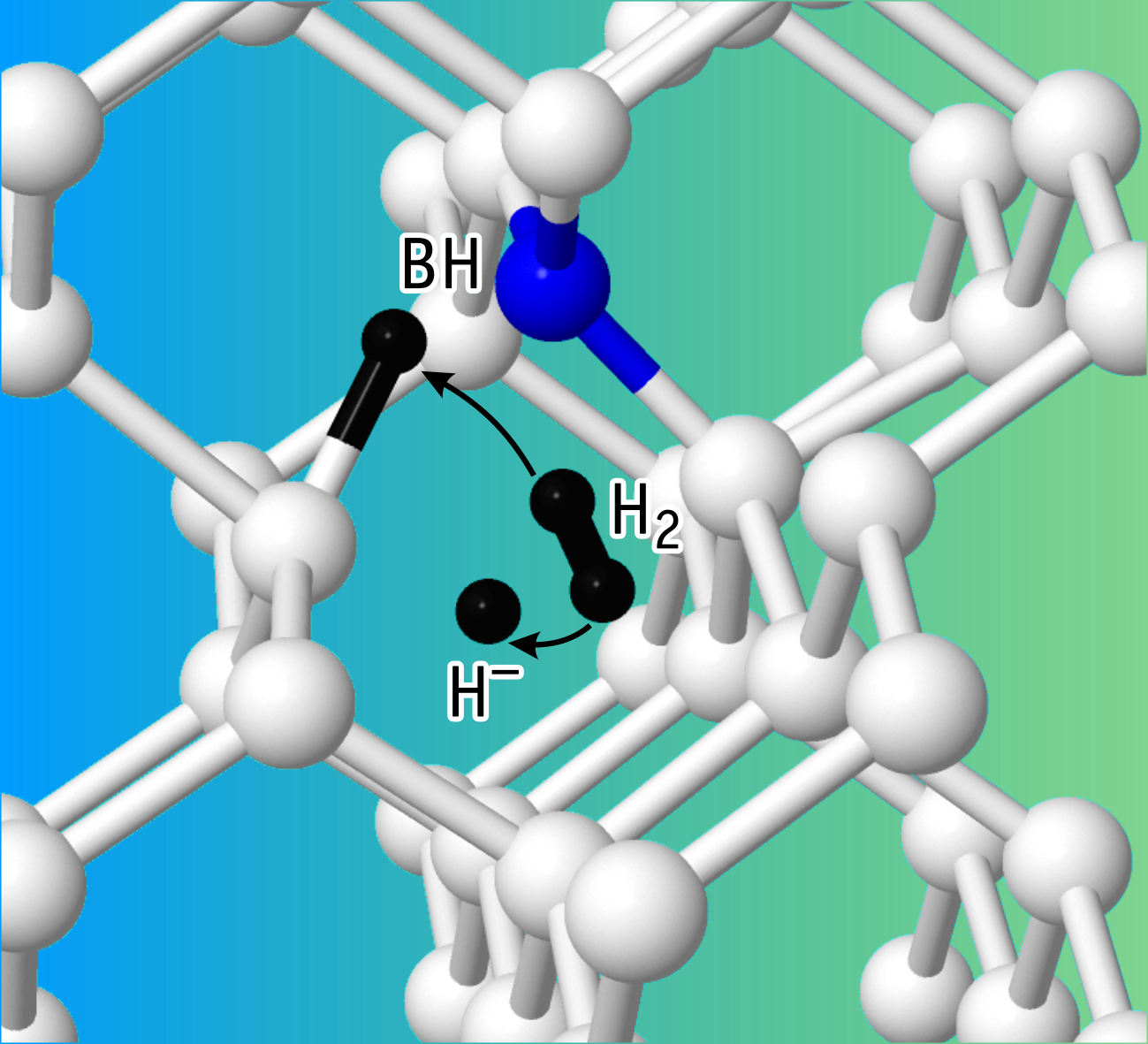}
\end{figure}

\noindent \textbf{Caption of ToC figure:} The thermodynamics of several
reactions involving atomic and molecular hydrogen with group-III acceptors
in silicon has been investigated theoretically. The results offer
a first-principles-level account of thermally- and carrier-activated
processes relevant to Light and elevated Temperature Induced Degradation
(LeTID) of Si-based solar cells.
\begin{acknowledgments}
We acknowledge the FCT through projects LA/P/0037/2020, UIDB/50025/2020,
UIDP/50025/2020 and 2021.09643.CPCA (Advanced Computing Project using
the Oblivion supercomputer). The work in the UK was funded by EPSRC
via grant EP/TO25131/1.
\end{acknowledgments}

\bibliographystyle{apsrev4-2}
%\bibliography{refs}
%apsrev4-2.bst 2019-01-14 (MD) hand-edited version of apsrev4-1.bst
%Control: key (0)
%Control: author (72) initials jnrlst
%Control: editor formatted (1) identically to author
%Control: production of article title (-1) disabled
%Control: page (0) single
%Control: year (1) truncated
%Control: production of eprint (0) enabled
%

\end{document}

\title{Supporting Information: Hydrogen reactions with dopants and impurities
in solar silicon from first principles}
\author{José Coutinho}
\affiliation{i3N and Department of Physics, University of Aveiro, Campus Santiago,
3810-193 Aveiro, Portugal}
\email{jose.coutinho@ua.pt}

\author{Diana Gomes}
\affiliation{i3N and Department of Physics, University of Aveiro, Campus Santiago,
3810-193 Aveiro, Portugal}
\author{Vitor J. B. Torres}
\affiliation{i3N and Department of Physics, University of Aveiro, Campus Santiago,
3810-193 Aveiro, Portugal}
\author{Tarek O. Abdul Fattah}
\affiliation{Photon Science Institute and Department of Electrical and Electronic
Engineering, The University of Manchester, Manchester M13 9PL, United
Kingdom}
\author{Vladimir P. Markevich}
\affiliation{Photon Science Institute and Department of Electrical and Electronic
Engineering, The University of Manchester, Manchester M13 9PL, United
Kingdom}
\author{Anthony R. Peaker}
\affiliation{Photon Science Institute and Department of Electrical and Electronic
Engineering, The University of Manchester, Manchester M13 9PL, United
Kingdom}
\maketitle

\noindent \emph{Published with Solar RRL 8, 2300639 (2024)}

\noindent DOI:\href{https://doi.org/10.1002/solr.202300639}{10.1002/solr.202300639}

\noindent 
\begin{figure}[H]
\includegraphics[width=7cm]{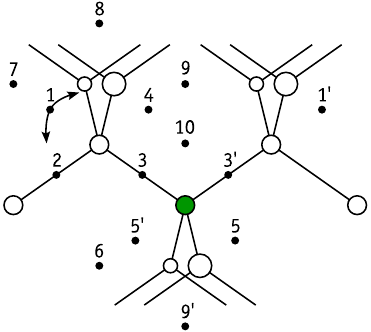}

Figure S1. Schematic geometries of the different PH pairs that were
investigated. The position of the H atom is indicated by a small dot
along with an index. Primed and non-primed indexes are equivalent
by symmetry. P and Si atoms are shown in green and white, respectively.
These correspond to approximate positions in the initial geometries
(before atomic relaxation). Structures that were not stable and ``collapsed''
to a lower energy configuration upon energy minimization are indicated
Table~S1 below. The two-way arrow indicates the wagging mode of H
discussed in the context of the light/injection enhanced dissociation
of PH.
\end{figure}

\noindent 
\begin{table}[H]
Table S1. Calculated relative energies of PH pairs with respect to
the lowest energy configuration (zero energy). The structure index
(first column) corresponds to the location of H as indicated in Figure~S1
above. Some structures have a $\textrm{PH}_{X}$ label as used in
the article. For some structures, $\rightarrow i$ indicates that
they are unstable in that specific charge state, and relax to another
structure with index $i$ during force and energy minimization. Configurations
without energy did not possess gap states capable of trapping holes.
No acceptor (negatively charged) states were found. All values are
in eV.

\begin{tabular}{llr@{\extracolsep{0pt}.}lr@{\extracolsep{0pt}.}lr@{\extracolsep{0pt}.}l}
\hline 
Index & Label & \multicolumn{2}{c}{$E_{\textrm{rel}}(q=0)$} & \multicolumn{2}{c}{$E_{\textrm{rel}}(q=+1)$} & \multicolumn{2}{c}{$E_{\textrm{rel}}(q=+2)$}\tabularnewline
\hline 
1 & $\textrm{PH}_{\textrm{AB}}$ & \multicolumn{2}{c}{\textbf{$\mathbf{0.00}$}} & \multicolumn{2}{c}{} & \multicolumn{2}{c}{}\tabularnewline
2 & $\textrm{PH}_{\textrm{BC1}}$ & \multicolumn{2}{c}{$1.14$} & \multicolumn{2}{c}{$0.14$} & \multicolumn{2}{c}{$0.31$}\tabularnewline
3 & $\textrm{PH}_{\textrm{BC}}$ & \multicolumn{2}{c}{$0.45$} & \multicolumn{2}{c}{$\mathbf{0.00}$} & \multicolumn{2}{c}{}\tabularnewline
4 &  & \multicolumn{2}{c}{$0.35$} & \multicolumn{2}{c}{$0.20$} & \multicolumn{2}{c}{}\tabularnewline
5 &  & \multicolumn{2}{c}{$0.53$} & \multicolumn{2}{c}{$0.31$} & \multicolumn{2}{c}{}\tabularnewline
6 &  & \multicolumn{2}{c}{$0.61$} & \multicolumn{2}{c}{$0.38$} & \multicolumn{2}{c}{}\tabularnewline
7 &  & \multicolumn{2}{c}{$\rightarrow1$} & \multicolumn{2}{c}{$0.27$} & \multicolumn{2}{c}{}\tabularnewline
8 &  & \multicolumn{2}{c}{$0.70$} & \multicolumn{2}{c}{$0.29$} & \multicolumn{2}{c}{}\tabularnewline
9 &  & \multicolumn{2}{c}{$0.39$} & \multicolumn{2}{c}{$0.37$} & \multicolumn{2}{c}{}\tabularnewline
10 &  & \multicolumn{2}{c}{$\rightarrow4$} & \multicolumn{2}{c}{$\rightarrow3$} & \multicolumn{2}{c}{}\tabularnewline
11 & $\textrm{P}+\textrm{H}_{\textrm{BC}}$ & \multicolumn{2}{c}{$1.36$} & \multicolumn{2}{c}{$0.14$} & \multicolumn{2}{c}{$\mathbf{0.00}$}\tabularnewline
\hline 
\end{tabular}
\end{table}

\noindent 
\begin{figure*}
\includegraphics[width=16cm]{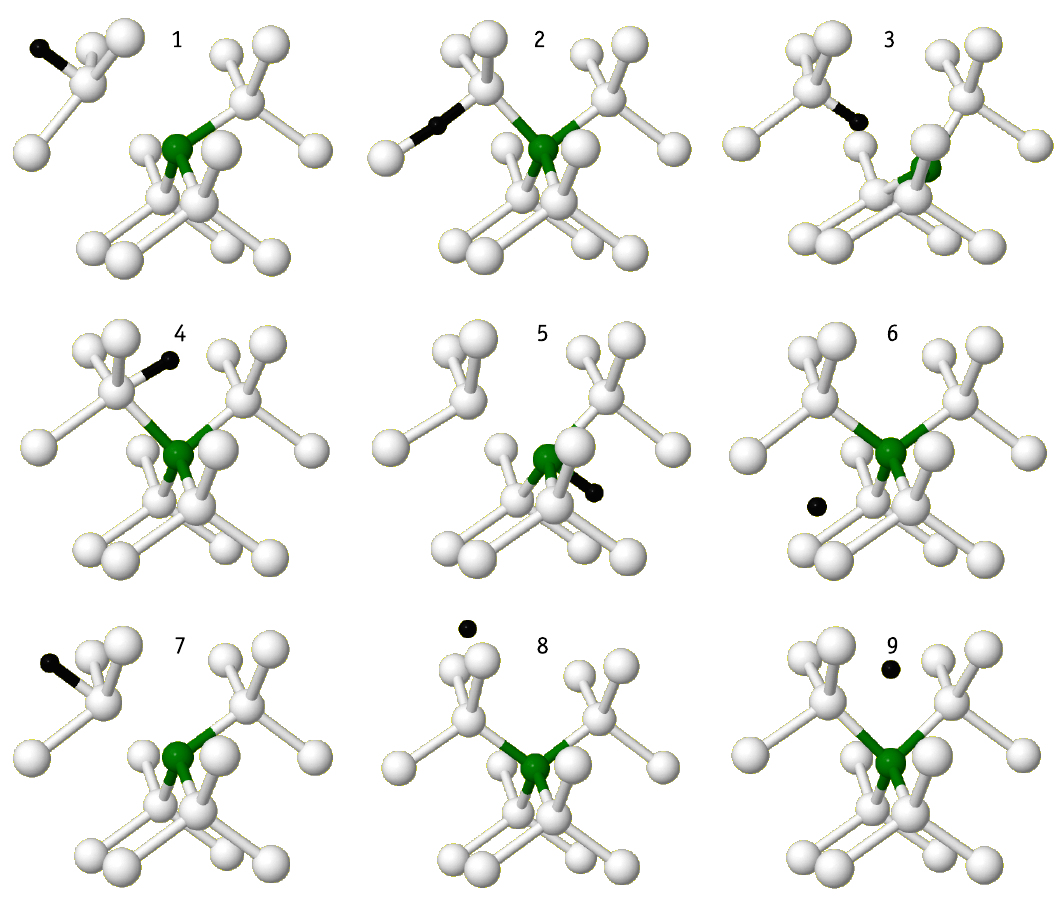}\\

\noindent \raggedright{}Figure S2. Atomistic geometries of several
PH pairs after relaxation. P, Si and H atoms are shown in green, white,
and black respectively. All structures but number 7 (which is positively
charged) correspond to the neutral charge state. Relative energies
are indicated in Table~S1.
\end{figure*}